\documentclass[prd,nofootinbib,twocolumn,showpacs,superscriptaddress]{revtex4}
\usepackage{graphics,rotating,multirow,bm,color,hyperref,amsmath,amssymb}
\hypersetup{
    colorlinks=true,
    linkcolor=green,
    citecolor=blue,
}

\begin{document}
\title{Structure Formation by the Fifth Force: Segregation of
Baryons and Dark Matter}
\author{Baojiu~Li}
\email[Email Address: ]{b.li@damtp.cam.ac.uk} \affiliation{Centre
for Theoretical Cosmology, DAMTP, Centre for Mathematical
Sciences, University of Cambridge, Wilberforce Road, Cambridge CB3
0WA, UK} \affiliation{Kavli Institute for Cosmology Cambridge,
Madingley Road, Cambridge CB3 0HA, UK}
\author{Hongsheng~Zhao}\email[Email Address: ]{hz4@st-andrews.ac.uk} \affiliation{SUPA,
University of St.~Andrews, North Haugh, Fife, KY16 9SS, UK}
\date{\today}

\begin{abstract}
In this paper we present the results of $N$-body simulations with
a scalar field coupled differently to cold dark matter (CDM) and
baryons. The scalar field potential and coupling function are
chosen such that the scalar field acquires a heavy mass in regions
with high CDM density and thus behaves like a chameleon. We focus
on how the existence of the scalar field affects the formation of
nonlinear large-scale structure, and how the different couplings
of the scalar field to baryons and CDM particles lead to different
distributions and evolutions for these two matter species, both on
large scales and inside virialized halos. As expected, the
baryon-CDM segregation increases in regions where the fifth force
is strong, and little segregation in dense regions. We also
introduce an approximation method to identify the virialized halos
in coupled scalar field models which takes into account the scalar
field coupling and which is easy to implement numerically. It is
find that the chameleon nature of the scalar field makes the
internal density profiles of halos dependent on the environment in
a very nontrivial way.
\end{abstract}

\maketitle

\section{Introduction}

\label{sect:intro}

The origin and nature of the dark energy \cite{Copeland2006} is
one of the most difficult challenges facing physicists and
cosmologists now. Among all the proposed models to tackle this
problem, a scalar field is perhaps the most popular one up to now.
The scalar field, denoted by $\varphi$, might only interact with
other matter species through gravity, or have a coupling to normal
matter and therefore producing a fifth force on matter particles.
This latter idea has seen a lot of interests in recent years, in
the light that such a coupling could potentially alleviate the
coincidence problem of dark energy \cite{Amendola2000} and that it
is commonly predicted by low energy effective theories from a
fundamental theory.

Nevertheless, if there is a coupling between the scalar field and
baryonic particles, then stringent experimental constraints might
be placed on the fifth force on the latter provided that the
scalar field mass is very light (which is needed for the dark
energy). Such constraints severely limit the viable parameter
space of the model. Different ways out of the problem have been
proposed, of which the simplest one is to have the scalar field
coupling to dark matter only but not to standard model particles,
therefore evading those constraints entirely. This is certainly
possible, especially because both dark matter and dark energy are
unknown to us and they may well have a common origin. Another
interesting possibility is to have the chameleon mechanism
\cite{Khoury2004a, Khoury2004b, Mota2006, Mota2007}, by virtue of
which the scalar field acquires a large mass in high density
regions and thus the fifth force becomes undetectablly
short-ranged, and so also evades the constraints.

Study of the cosmological effect of a chameleon scalar field shows
that the fifth force is so short-ranged that it has negligible
effect in the large scale structure formation \cite{Brax2004} for
certain choices of the scalar field potential. But it is possible
that the scalar field has a large enough mass in the solar system
to pass any constraints, and at the same time has a low enough
mass (thus long range forces) on cosmological scales, producing
interesting phenomenon in the structure formation. This is the
case of some $f(R)$ gravity models \cite{Carroll2004, Nojiri2003},
which survives solar system tests thanks again to the chameleon
effect \cite{Navarro2007, Li2007, Hu2007, Brax2008}. Note that the
$f(R)$ gravity model is mathematically equivalent to a scalar
field model with matter coupling.

No matter whether the scalar field couples with dark matter only
or with all matter species, it is of general interests to study
its effects in cosmology, especially in the large scale structure
formation. Indeed, at the linear perturbation level there have
been a lot of studies about the coupled scalar field and $f(R)$
gravity models which enable us to have a much clearer picture
about their behaviors now. But linear perturbation studies do not
conclude the whole story, because it is well known that the matter
distribution at late times becomes nonlinear, making the behavior
of the scalar field more complex and the linear analysis
insufficient to produce accurate results to confront with
observations. For the latter purpose the best way is to perform
full $N$-body simulations \cite{Bertschinger1998} to evolve the
individual particles step by step.

$N$-body simulations for scalar field and relevant models have
been performed before \cite{Linder2003, Mainini2003, Springel2007,
Kesden2006, Farrar2007, Keselman2009, Maccio2004, Baldi2008}. For
example, in \cite{Maccio2004} the simulation is about a specific
coupled scalar field model. This study however does not obtain a
full solution to the spatial configuration of the scalar field,
but instead simplifies the simulation by assuming that the scalar
field's effect is to change the value of the gravitational
constant, and presenting an justifying argument for such an
approximation. As discussed in \cite{Li2009, Zhao2009}, this
approximation is only good in certain parameter spaces and for
certain choices of the scalar field potential, and therefore full
simulations tare needed to study the scalar field behaviour more
rigorously.

Recently there have also appeared $N$-body simulations of the
$f(R)$ gravity model \cite{Oyaizu2008, Oyaizu2008b}, which do
solve the scalar degree of freedom explicitly. However, embedded
in the $f(R)$ framework there are some limitations in the
generality of these works. As a first thing, $f(R)$ gravity model
(no matter what the form $f$ is) only corresponds to the couple
scalar field models for a specific value of coupling strength
\cite{Amendola2007}. Second, in $f(R)$ models the correction to
standard general relativity is through the modification to the
Poisson equation and thus to the gravitational potential as a
whole \cite{Oyaizu2008}, while in the coupled scalar field models
we could clearly separate the scalar fifth force from gravity and
analyze the former directly \cite{Li2009}. Also, in $f(R)$ models,
as well as the scalar-tensor theories, the coupling between matter
and the scalar field is universal (the same to dark matter and
baryons), while in the couple scalar field models it is
straightforward to switch on/off the coupling to baryons and study
the effects on baryonic and dark matter clusterings respectively
(as we will do in this paper). Correspondingly, the general
framework of $N$-body simulations in coupled scalar field models
could also handle the situation where the chameleon effect is
absent and/or scalar field only couples to dark matter, and thus
provide a testbed for possible violations of the weak equivalence
principle.

In this paper we shall go beyond \cite{Li2009} and consider the
case where the chameleon scalar field couples differently to
different species of matter. To be explicit, we consider two
matter species, and let one of them have no coupling to the scalar
field. Because it is commonly believed that normal baryons, being
observable in a variety of experiments, should have extremely weak
(if any) coupling to scalar fields, we call the uncoupled matter
species in our simulation "baryons". It is however reminded here
that this matter species is not really baryonic in the sense that
it does not experience normal baryonic interactions. The inclusion
of true baryons will make the investigation more complicated and
is thus beyond the scope of the present work.

The paper is organized as follows: in \S~\ref{sect:eqns} we list
the essential equations to be implemented in the $N$-body
simulations and describe briefly the difference from normal LCDM
simulations. \S~\ref{sect:simu} is the main body of the paper, in
which \S~\ref{subsect:simu_detail} gives the details about our
simulations, such as code description and parameter set-up,
\S~\ref{subsect:simu_result} displays some preliminary results for
visualization, such as baryon/CDM distribution, potential/scalar
field configuration and the correlation between the fifth force
(for CDM particles) and gravity; \S~\ref{subsect:simu_ps}
quantifies the nonlinear matter power spectrum of our model,
especially the difference from LCDM results and the bias between
CDM and baryons; \S~\ref{subsect:simu_mf} briefly describes the
essential modifications one must bear in mind when identifying
\emph{virialized} halos from the simulation outputs and shows the
mass functions for our models; \S~\ref{subsect:simu_prof} we pick
out two halos from our simulation box and analyzes their total
internal profiles, as well as their baryonic/CDM density profiles.
We finally summarize in \S~\ref{sect:con}.

\section{The Equations}

\label{sect:eqns}

In this section we first describe the method to simulate structure
formation with two differently coupled matter species and the
appropriate equations to be used. Those equations for a single
matter species have been discussed in details previously in
\cite{Li2009}, but the inclusion of different matter species
requires further modifications and we list all these for
completeness.

\subsection{The Fundamental Equations}

\label{subsect:eqn_original}

The Lagrangian for our coupled scalar field model is
\begin{eqnarray}\label{eq:Lagrangian}
\mathcal{L} &=&
\frac{1}{2}\left[\frac{R}{\kappa}-\nabla^{a}\varphi\nabla_{a}\varphi\right]
+ V(\varphi) - C(\varphi)\mathcal{L}_{\mathrm{CDM}} +
\mathcal{L}_{\mathrm{S}}\ \
\end{eqnarray}
where $R$ is the Ricci scalar, $\kappa=8\pi G$ with $G$ Newton's
constant, $\varphi$ is the scalar field, $V(\varphi)$ is its
potential energy and $C(\varphi)$ its coupling to dark matter,
which is assumed to be cold and described by the Lagrangian
$\mathcal{L}_{\mathrm{CDM}}$. $\mathcal{L}_{\mathrm{S}}$ includes
all other matter species, in particular our \emph{baryons}. The
contribution from photons and neutrinos in the $N$-body
simulations (for late times, \emph{i.e.}, $z\sim\mathcal{O}(1)$)
is negligible, but should be included when generating the matter
power spectrum from which the initial conditions for our $N$-body
simulations are obtained (see below).

The dark matter Lagrangian for a point-like particle with bare
mass $m_{0}$ is
\begin{eqnarray}\label{eq:DMLagrangian}
\mathcal{L}_{\mathrm{CDM}}(\mathbf{y}) &=&
-\frac{m_{0}}{\sqrt{-g}}\delta(\mathbf{y}-\mathbf{x}_{0})\sqrt{g_{ab}\dot{x}^{a}_{0}\dot{x}^{b}_{0}}
\end{eqnarray}
where $\mathbf{y}$ is the coordinate and $\mathbf{x}_{0}$ is the
coordinate of the centre of the particle. From this equation it
can be easily derived that
\begin{eqnarray}\label{eq:DMEMT_particle}
T^{ab}_{\mathrm{CDM}} &=&
\frac{m_{0}}{\sqrt{-g}}\delta(\mathbf{y}-\mathbf{x}_{0})
\dot{x}^{a}_{0}\dot{x}^{b}_{0}.
\end{eqnarray}
Also, because $g_{ab}\dot{x}^{a}_{0}\dot{x}^{b}_{0}\equiv
g_{ab}u^{a}u^{b}=1$ where $u^{a}$ is the four velocity of the dark
matter particle, the Lagrangian could be rewritten as
\begin{eqnarray}\label{eq:DMLagrangian2}
\mathcal{L}_{\mathrm{CDM}}(\mathbf{y}) &=&
-\frac{m_{0}}{\sqrt{-g}}\delta(\mathbf{y}-\mathbf{x}_{0}),
\end{eqnarray}
which will be used below.

Eq.~(\ref{eq:DMEMT_particle}) is just the energy momentum tensor
for a single dark matter particle. For a fluid with many particles
the energy momentum tensor will be
\begin{eqnarray}\label{eq:DMEMT_fluid}
T^{ab}_{\mathrm{CDM}} &=& \frac{1}{V}\int_{V}d^{4}y\sqrt{-g}
\frac{m_{0}}{\sqrt{-g}}\delta(y-x_{0})
\dot{x}^{a}_{0}\dot{x}^{b}_{0}\nonumber\\
&=& \rho_{\mathrm{CDM}}u^{a}u^{b},
\end{eqnarray}
in which $V$ is a volume microscopically large and macroscopically
small, and we have extended the 3-dimensional $\delta$ function to
a 4-dimensional one by adding a time component. Here $u^{a}$ is
the averaged 4-velocity of the collection of particles inside this
volume, and is not necessarily the same as the $4$-velocity of the
observer.

Meanwhile, using
\begin{eqnarray}
T^{ab} &=&
-\frac{2}{\sqrt{-g}}\frac{\delta\left(\sqrt{-g}\mathcal{L}\right)}{\delta
g_{ab}}
\end{eqnarray}
it is straightforward to show that the energy momentum tensor for
the scalar field is given by
\begin{eqnarray}\label{eq:phiEMT}
T^{\varphi ab} &=& \nabla^{a}\varphi\nabla^{b}\varphi -
g^{ab}\left[\frac{1}{2}\nabla_{c}\varphi\nabla^{c}\varphi -
V(\varphi)\right].
\end{eqnarray}
So the total energy momentum tensor is
\begin{eqnarray}\label{eq:EMT_tot}
T_{ab} &=& \nabla_{a}\varphi\nabla_{b}\varphi -
g_{ab}\left[\frac{1}{2}\nabla_{c}\varphi\nabla^{c}\varphi -
V(\varphi)\right]\nonumber\\
&& + C(\varphi)T^{\mathrm{CDM}}_{ab} + T^{\mathrm{S}}_{ab}
\end{eqnarray}
where $T^{\mathrm{CDM}}_{ab}=\rho_{\mathrm{CDM}}u_{a}u_{b}$,
$T^{\mathrm{S}}_{ab}$ is the energy momentum tensor for all other
matter species including baryons, and the Einstein equation is
\begin{eqnarray}\label{eq:EinsteinEq}
G_{ab} &=& \kappa T_{ab}
\end{eqnarray}
where $G_{ab}$ is the Einstein tensor. Note that due to the
coupling between the scalar field $\varphi$ and the dark matter,
the energy momentum tensors for either will not be conserved, and
we have
\begin{eqnarray}\label{eq:DM_energy_conservation}
\nabla_{b}T^{\mathrm{CDM}_{}ab} &=&
-\frac{C_{\varphi}(\varphi)}{C(\varphi)}\left(g^{ab}\mathcal{L}_{\mathrm{CDM}}
+ T^{\mathrm{CDM}ab}\right)\nabla_{b}\varphi\ \
\end{eqnarray}
where throughout this paper we shall use a $_{\varphi}$ to denote
the derivative with respect to $\varphi$.

Finally, the scalar field equation of motion (EOM) from the given
Lagrangian is
\begin{eqnarray}
\square\varphi + \frac{\partial V(\varphi)}{\partial\varphi} &=&
\frac{\partial
C(\varphi)}{\partial\varphi}\mathcal{L}_{\mathrm{CDM}}\nonumber
\end{eqnarray}
where $\square = \nabla^{a}\nabla_{a}$. Using
Eq.~(\ref{eq:DMLagrangian2}) it can be rewritten as
\begin{eqnarray}\label{eq:phiEOM}
\square\varphi + \frac{\partial V(\varphi)}{\partial\varphi} +
\rho_{\mathrm{CDM}}\frac{\partial C(\varphi)}{\partial\varphi} &=&
0.
\end{eqnarray}

Eqs.~(\ref{eq:EMT_tot}, \ref{eq:EinsteinEq},
\ref{eq:DM_energy_conservation}, \ref{eq:phiEOM}) summarize all
the physics that will be used in our analysis.

We will consider a special form for the scalar field potential,
\begin{eqnarray}\label{eq:potential}
V(\varphi) &=& \frac{V_{0}}
{\left[1-\exp\left(\beta\sqrt{\kappa}\varphi\right)\right]^{\mu}},
\end{eqnarray}
where $\mu$ and $\beta$ are dimensionless constants while $V_{0}$
has mass dimension four. As has been discussed in \cite{Li2009},
$\mu\ll1$ to evade observational constraints and $\beta$ can be
set to $-1$ without loss of generality, since we can always
rescale $\varphi$ as we wish. Meanwhile, the coupling between the
scalar field and dark matter particle is chosen as
\begin{eqnarray}\label{eq:coupling_function}
C(\varphi) &=& \exp(\gamma\sqrt{\kappa}\varphi),
\end{eqnarray}
where $\gamma>0$ is yet another dimensionless constant
characterizing the strength of the coupling.

As discussed in \cite{Li2009}, the two dimensionless parameters
$\mu$ and $\gamma$ have clear physical meanings: roughly speaking,
$\mu$ controls the time when the scalar field becomes important in
cosmology while $\gamma$ determines how important the scalar field
would ultimately be. In fact, the potential given in
Eq.~(\ref{eq:potential}) is partly motivated by the $f(R)$
cosmology \cite{Li2007}, in which the extra degree of freedom
behaves as a coupled scalar field in the Einstein frame. As we can
see from Eq.~(\ref{eq:potential}), the potential
$V\rightarrow\infty$ when $\varphi\rightarrow0$ while
$V\rightarrow V_{0}$ when $\varphi\rightarrow\infty$. In the
latter case, however, $C\rightarrow\infty$, so that the effective
total potential
\begin{eqnarray}
V_{eff}(\varphi) &=& V(\varphi) + \rho_{\mathrm{CDM}}C(\varphi)
\end{eqnarray}
has a global minimum at some finite $\varphi$. If the total
potential $V_{eff}(\varphi)$ is steep enough around this minimum,
then the scalar field becomes very heavy and thus follows its
minimum dynamically, as is in the case of the chameleon cosmology
(see \emph{e.g.}~\cite{Brax2004}). If $V_{eff}$ is not steep
enough at the minimum, however, the scalar field will experience a
more complicated evolution. These two different cases can be
obtained by choosing appropriate values of $\gamma$ and $\mu$: if
$\gamma$ is very large or $\mu$ is small then we run into the
former situation and if $\gamma$ is small and $\mu$ is large we
have the second. In reality, the situation can get even more
complicated because when $\gamma$, which characterizes the
coupling strength, increases, the CDM evolution could also get
severely affected, which in turn has back-reactions on the scalar
field itself.

\subsection{Nonrelativistic Limit}

The $N$-body simulation only probes the motion of particles at
late times, and we are not interested in extreme conditions such
as black hole formation/evolution, which mean that taking the
non-relativistic limit of the above equations should be a
sufficient approximation for our purpose.

The existence of the scalar field and its (different) couplings to
matter particles lead to the following changes to the $\Lambda$CDM
model: Firstly, the energy momentum tensor has a new piece of
contribution from the scalar field; secondly, the energy density
of dark matter in gravitational field equations is multiplied by
the function $C(\varphi)$, which is because the coupling to scalar
field essentially renormalizes the mass of dark matter particles;
thirdly, dark matter particles will not follow geodesics in their
motions as in $\Lambda$CDM, but rather the total force on them has
a contribution, the fifth force, from the exchange of scalar field
quanta; finally, CDM particles must be distinguished from baryons
so that the fifth force only acts on the former and these two
species only interact gravitationally. This last point is one main
difference between the present work and a previous one
\cite{Li2009}.

These imply that the following things need to be modified or
added:
\begin{enumerate}
    \item The scalar field $\varphi$ equation of motion, which determines
    the value of the scalar field at any given time and position;
    \item The Poisson equation, which determines the gravitational
    potential (and thus gravity) at any given time and position,
    according to the local energy density and pressure, which include
    the contribution from the scalar field (as obtained from $\varphi$ equation of
    motion);
    \item The total force on the dark matter particles, which is
    determined by the spatial configuration of $\varphi$,
    just like gravity is determined by the spatial configuration
    of the gravitational potential;
    \item The CDM and baryonic particles must be tagged
    respectively so that the code knows to assign forces correctly
    to different species.
\end{enumerate}
We shall describe these one by one now.

For the scalar field equation of motion, we denote $\bar{\varphi}$
as the background value of $\varphi$ and
$\delta\varphi\equiv\varphi-\bar{\varphi}$ as the scalar field
perturbation. Then Eq.~(\ref{eq:phiEOM}) could be rewritten as
\begin{eqnarray}
\ddot{\delta\varphi} + 3H\dot{\delta\varphi} +
\vec{\nabla}_{\mathbf{r}}^{2}\varphi +
V_{,\varphi}(\varphi)-V_{,\varphi}(\bar{\varphi})\nonumber\\ +
\rho_{\mathrm{CDM}}C_{,\varphi}(\varphi) -
\bar{\rho}_{\mathrm{CDM}}C_{,\varphi}(\bar{\varphi}) &=&
0\nonumber
\end{eqnarray}
by subtracting the corresponding background equation from it. Here
$\vec{\nabla}_{\mathbf{r}a}$ is the covariant spatial derivative
with respect to the physical coordinate $\mathbf{r}=a\mathbf{x}$
with $\mathbf{x}$ the conformal coordinate, and
$\vec{\nabla}^{2}_{\mathbf{r}}=\vec{\nabla}_{\mathbf{r}a}\vec{\nabla}_{\mathbf{r}}^{a}$.
$\vec{\nabla}_{\mathbf{r}a}$ is essentially the
$\hat{\nabla}_{a}$, but because here we are working in the weak
field limit we approximate it as $\vec{\nabla}_{\mathbf{r}}^{2} =
-\left(\partial^{2}_{r_{x}}+\partial^{2}_{r_{y}}+\partial^{2}_{r_{z}}\right)$
by assuming a flat background; the minus sign is because our
metric convention is $(+,-,-,-)$ instead of $(-,+,+,+)$. For the
simulation here we will also work in the quasi-static limit,
assuming that the spatial gradient is much larger than the time
derivative,
$|\vec{\nabla}_{\mathbf{r}}\varphi|\gg|\frac{\partial\varphi}{\partial
t}|$ (which will be justified below). Thus the above equation can
be further simplified as
\begin{eqnarray}\label{eq:WFphiEOM}
&&c^{2}\partial_{\mathbf{x}}^{2}(a\delta\varphi)\\ &=&
a^{3}\left[V_{,\varphi}(\varphi)-V_{,\varphi}(\bar{\varphi}) +
\rho_{\mathrm{CDM}}C_{,\varphi}(\varphi) -
\bar{\rho}_{\mathrm{CDM}}C_{,\varphi}(\bar{\varphi})\right],\nonumber
\end{eqnarray}
in which $\partial^{2}_{\mathbf{x}}=-\vec{\nabla}_{\mathbf{x}}^{2}
=
+\left(\partial^{2}_{x}+\partial^{2}_{y}+\partial^{2}_{z}\right)$
is with respect to the conformal coordinate $\mathbf{x}$ so that
$\vec{\nabla}_{\mathbf{x}}=a\vec{\nabla}_{\mathbf{r}}$, and we
have restored the factor $c^{2}$ in front of
$\vec{\nabla}_{\mathbf{x}}^{2}$ (the $\varphi$ here and in the
remaining of this paper is $c^{-2}$ times the $\varphi$ in the
original Lagrangian unless otherwise stated). Note that here $V$
and $\rho_{\mathrm{CDM}}$ both have the dimension of\emph{ mass
}density rather than \emph{energy} density.

Next look at the Poisson equation, which is obtained from the
Einstein equation in weak-field and slow-motion limits. Here the
metric could be written as
\begin{eqnarray}
ds^{2} &=& (1+2\phi)dt^{2} - (1-2\psi)\delta_{ij}dr^{i}dr^{j}
\end{eqnarray}
from which we find that the time-time component of the Ricci
curvature tensor $R^{0}_{\ 0}=-\vec{\nabla}_{\mathbf{r}}^{2}\phi$,
and then the Einstein equation
$R_{ab}=\kappa\left(T_{ab}-\frac{1}{2}g_{ab}T\right)$ gives
\begin{eqnarray}\label{eq:EinsteinEqn}
R^{0}_{\ 0}\ =\ -\vec{\nabla}_{\mathbf{r}}^{2}\phi\ =\
\frac{\kappa}{2}(\rho_{\mathrm{TOT}}+3p_{\mathrm{TOT}})
\end{eqnarray}
where $\rho_{\mathrm{TOT}}$ and $p_{\mathrm{TOT}}$ are
respectively the total energy density and pressure. The quantity
$\vec{\nabla}_{\mathbf{r}}^{2}\phi$ can be expressed in terms of
the comoving coordinate $\mathbf{x}$ as
\begin{eqnarray}
\vec{\nabla}_{\mathbf{r}}^{2}\phi &=& \frac{1}{a^{2}}
\vec{\nabla}_{\mathbf{x}}^{2}\left(\frac{\Phi}{a} -
\frac{1}{2}a\ddot{a}\mathbf{x}^{2}\right)\nonumber\\
&=& \frac{1}{a^{3}}\vec{\nabla}_{\mathbf{x}}^{2}\Phi -
3\frac{\ddot{a}}{a}
\end{eqnarray}
where we have defined a new Newtonian potential
\begin{eqnarray}\label{eq:newphi}
\Phi &\equiv& a\phi + \frac{1}{2}a^{2}\ddot{a}\mathbf{x}^{2}
\end{eqnarray}
and used $\vec{\nabla}_{\mathbf{x}}^{2}\mathbf{x}^{2}=6$. Thus
\begin{eqnarray}
\vec{\nabla}_{\mathbf{x}}^{2}\Phi &=&
a^{3}\left(\vec{\nabla}_{\mathbf{r}}^{2}\phi +
3\frac{\ddot{a}}{a}\right)\\
&=& -
a^{3}\left[\frac{\kappa}{2}(\rho_{\mathrm{TOT}}+3p_{\mathrm{TOT}})
-
\frac{\kappa}{2}(\bar{\rho}_{\mathrm{TOT}}+3\bar{p}_{\mathrm{TOT}})\right]\nonumber
\end{eqnarray}
where in the second step we have used Eq.~(\ref{eq:EinsteinEqn})
and the Raychaudhrui equation, and an overbar means the background
value of a quantity. Because the energy momentum tensor for the
scalar field is given by Eq.~(\ref{eq:phiEMT}), it is easy to show
that $\rho^{\varphi}+3p^{\varphi} =
2\left[\dot{\varphi}^{2}-V(\varphi)\right]$ and so
\begin{eqnarray}
&&\vec{\nabla}_{\mathbf{x}}^{2}\Phi\nonumber\\
&=& -4\pi
Ga^{3}\left\{\rho_{\mathrm{CDM}}C(\varphi)+\rho_{\mathrm{B}} +
2\left[\dot{\varphi}^{2}-V(\varphi)\right]\right\}\nonumber\\
&& +4\pi Ga^{3} \left\{\bar{\rho}_{\mathrm{CDM}}C(\bar{\varphi})+
\bar{\rho}_{\mathrm{B}}+
2\left[\dot{\bar{\varphi}}^{2}-V(\bar{\varphi})\right]\right\}.\nonumber
\end{eqnarray}
Now in this equation $\dot{\varphi}^{2} - \dot{\bar{\varphi}}^{2}
= 2\dot{\varphi}\dot{\delta\varphi} + \dot{\delta\varphi}^{2} \ll
(\vec{\nabla}_{\mathbf{r}}\varphi)^{2}$ in the quasi-static limit
and so could be dropped safely. So we finally have
\begin{eqnarray}\label{eq:WFPoisson}
\partial_{\mathbf{x}}^{2}\Phi &=& 4\pi Ga^{3}
\left[\rho_{\mathrm{CDM}}C(\varphi)-\bar{\rho}_{\mathrm{CDM}}C(\bar{\varphi})\right]\nonumber\\
&&+4\pi Ga^{3}
\left[\rho_{\mathrm{B}}-\bar{\rho}_{\mathrm{B}}\right] - 8\pi
Ga^{3}\left[V(\varphi)-V(\bar{\varphi})\right].\ \
\end{eqnarray}

Finally, for the equation of motion of the dark matter particle,
consider Eq.~(\ref{eq:DM_energy_conservation}). Using
Eqs.~(\ref{eq:DMEMT_particle}, \ref{eq:DMLagrangian2}), this can
be reduced to
\begin{eqnarray}\label{eq:DMEOM}
\ddot{x}^{a}_{0} + \Gamma^{a}_{bc}\dot{x}^{b}_{0}\dot{x}^{c}_{0}
&=&
\left(g^{ab}-u^{a}u^{b}\right)\frac{C_{\varphi}(\varphi)}{C(\varphi)}\nabla_{b}\varphi.
\end{eqnarray}
Obviously the left hand side is the conventional geodesic equation
and the right hand side is the new fifth force due to the coupling
to the scalar field. Note that because $g^{ab}-u^{a}u^{b}=h^{ab}$
is the projection tensor that projects any 4-tensor into the
3-space perpendicular to $u^{a}$, so
$\left(g^{ab}-u^{a}u^{b}\right)\nabla_{a}=\hat{\nabla}^{b}$ is the
spatial derivative in the 3-space of the observer and
perpendicular to $u^a$; consequently the fifth force
$\frac{C_{\varphi}(\varphi)}{C(\varphi)}\hat{\nabla}_{a}\varphi=\hat{\nabla}_{a}\log
C(\varphi)$ has no component parallel to $u^a$ (the time
component), indicating that the energy density of CDM will be
conserved and only the particle trajectories are modified, as
mentioned in \cite{Li2009}. Remember that $u^a$ in
Eq.~(\ref{eq:DMEOM}) is the 4-velocity of individual particles,
but from Eq.~(\ref{eq:WFphiEOM}) we see that $\delta\varphi$ is
computed in the fundamental observer's frame (where density
perturbation is calculated), so if we also want to work on
Eq.~(\ref{eq:DMEOM}) in the fundamental observer's frame (so that
we can use the $\delta\varphi$ from Eq.~(\ref{eq:WFphiEOM})
directly), then we must rewrite Eq.~(\ref{eq:DMEOM}) by
substituting
\begin{eqnarray}
\left(g^{ab}-u^{a}u^{b}\right)\nabla_{b}\varphi &=&
\left(g^{ab}-\tilde{u}^{a}\tilde{u}^{b}-\tilde{u}^{a}v^{b}-\tilde{u}^{b}v^{a}\right)\nabla_{b}\varphi\nonumber\\
&\approx&
\left(g^{ab}-\tilde{u}^{a}\tilde{u}^{b}\right)\nabla_{b}\varphi -
\dot{\varphi}v^{a}\nonumber
\end{eqnarray}
up to first order in perturbations, in which $\tilde{u}^a$ is the
4-velocity of the fundamental observer and $v$ is the peculiar
velocity of the particle. Then the first term in the above
expression is the gradient of $\delta\varphi$ observed by the
fundamental observer (rather than an observer comoving with the
particle) and the second term is a velocity dependent acceleration
\cite{Maccio2004}. In \cite{Baldi2008} it is claimed that the
second term is of big importance; in our simulations, however,
this term will be neglected (from here on) because it depends on
$\dot{\varphi}$, which is very small due to the chameleon nature
of the model. We have checked in a linear perturbation computation
that removing this term only changes the matter power spectrum by
less than 0.0001\%.

Now in the non-relativistic limit the spatial components of
Eq.~(\ref{eq:DMEOM}) can be written as
\begin{eqnarray}\label{eq:DMEOM_physical}
\frac{d^{2}\mathbf{r}}{dt^{2}} &=& -\vec{\nabla}_{\mathbf{r}}\phi
-
\frac{C_{\varphi}(\varphi)}{C(\varphi)}\vec{\nabla}_{\mathbf{r}}\varphi
\end{eqnarray}
where $t$ is the physical time coordinate. If we instead use the
comoving coordinate $\mathbf{x}$, then this becomes
\begin{eqnarray}
\ddot{\mathbf{x}} + 2\frac{\dot{a}}{a}\dot{\mathbf{x}} &=&
-\frac{1}{a^{3}}\vec{\nabla}_{\mathbf{x}}\Phi  -
\frac{1}{a^{2}}\frac{C_{\varphi}(\varphi)}{C(\varphi)}
\vec{\nabla}_{\mathbf{x}}\varphi
\end{eqnarray}
where we have used Eq.~(\ref{eq:newphi}). The canonical momentum
conjugate to $\mathbf{x}$ is $\mathbf{p}=a^{2}\dot{\mathbf{x}}$ so
we have now
\begin{eqnarray}\label{eq:WFdxdtcomov}
\frac{d\mathbf{x}}{dt} &=& \frac{\mathbf{p}}{a^{2}},\\
\label{eq:WFdpdtcomov} \frac{d\mathbf{p}_{\mathrm{CDM}}}{dt} &=&
-\frac{1}{a}\vec{\nabla}_{\mathbf{x}}\Phi -
\frac{C_{\varphi}(\varphi)}{C(\varphi)}\vec{\nabla}_{\mathbf{x}}\varphi,\\
\label{eq:WFdpdtcomovb} \frac{d\mathbf{p}_{\mathrm{B}}}{dt} &=&
-\frac{1}{a}\vec{\nabla}_{\mathbf{x}}\Phi,
\end{eqnarray}
in which Eq.~(\ref{eq:WFdpdtcomov}) is for CDM particle and
Eq.~(\ref{eq:WFdpdtcomovb}) is for baryons. Note that according to
Eq.~(\ref{eq:WFdpdtcomov}) the quantity $a\log[C(\varphi)]$ acts
as a new piece of potential: the potential for the fifth force.
This is an important observation and we will come back to it later
when we calculate the escape velocity of CDM particles within a
virialized halo.

Eqs.~(\ref{eq:WFphiEOM}, \ref{eq:WFPoisson}, \ref{eq:WFdxdtcomov},
\ref{eq:WFdpdtcomov}, \ref{eq:WFdpdtcomovb}) will be used in the
code to evaluate the forces on the dark matter particles and
evolve their positions and momenta in time.

\subsection{Internal Units}

\label{subsect:eqn_unit}

In our numerical simulation we use a modified version of MLAPM
(\cite{MLAPM}, see \ref{subsect:simu_detail}), and we will have to
change our above equations in accordance with the internal units
used in that code. Here we briefly summarize the main features.

MLAPM code uses the following internal units (with subscript
$_{c}$):
\begin{eqnarray}
\mathbf{x}_{c} &=& \mathbf{x}/B,\nonumber\\
\mathbf{p}_{c} &=& \mathbf{p}/(H_{0}B)\nonumber\\
t_{c} &=& tH_{0}\nonumber\\
\Phi_{c} &=& \Phi/(H_{0}B)^{2}\nonumber\\
\rho_{c} &=& \rho/\bar{\rho},
\end{eqnarray}
in which $B$ is the present size of the simulation box and $H_{0}$
is the present Hubble constant, and $\rho$, with subscript, can
represent the density for either CDM ($\rho_{c,\mathrm{CDM}}$) or
baryons ($\rho_{c,\mathrm{B}}$). Using these newly-defined
quantities, it is easy to check that Eqs.~(\ref{eq:WFdxdtcomov},
\ref{eq:WFdpdtcomov}, \ref{eq:WFPoisson}, \ref{eq:WFphiEOM}) could
be rewritten as
\begin{eqnarray}\label{eq:INTdxdtcomov}
\frac{d\mathbf{x}_{c}}{dt_{c}} &=& \frac{\mathbf{p}_{c}}{a^{2}},\\
\label{eq:INTdpdtcomov} \frac{d\mathbf{p}_{c}}{dt_{c}} &=&
-\frac{1}{a}\nabla\Phi_{c} \left[-\frac{C_{,\varphi}}{C}c^{2}\nabla\varphi\right],\\
\label{eq:INTPoisson}\nabla^{2}\Phi_{c} &=&
\frac{3}{2}\Omega_{\mathrm{CDM}}\bar{C}
\left(\rho_{c,\mathrm{CDM}}\frac{C}{\bar{C}}-1\right)\nonumber\\
&&+\frac{3}{2}\Omega_{\mathrm{B}}\left(\rho_{c,\mathrm{B}}-1\right)
- \kappa\frac{V-\bar{V}}{H^{2}_{0}}a^{3},
\end{eqnarray}
and
\begin{eqnarray}\label{eq:INTphiEOM}
&&\frac{c^{2}}{\left(BH_{0}\right)^{2}}\nabla^{2}\left(a\varphi\right)\nonumber\\
&=&
\frac{3}{\kappa}\Omega_{\mathrm{CDM}}\bar{C}_{,\varphi}\left(\rho_{c,\mathrm{CDM}}\frac{C_{,\varphi}}{\bar{C}_{,\varphi}}-1\right)
+ \frac{V_{,\varphi}-\bar{V}_{,\varphi}}{H^{2}_{0}}a^{3},
\end{eqnarray}
where $\Omega_{\mathrm{CDM}}$ is the present CDM fractional energy
density, we have again restored the factor $c^{2}$ and again the
$\varphi$ is $c^{-2}$ times the $\varphi$ in the original
Lagrangian. Note that in Eq.~(\ref{eq:INTdpdtcomov}) the term in
the bracket on the right hand side only applies to CDM but not to
baryons. Also note that from here on we shall use
$\nabla\equiv\vec{\partial}_{\mathbf{x}_{c}},
\nabla^{2}\equiv\vec{\partial}_{\mathbf{x}_{c}}\cdot\vec{\partial}_{\mathbf{x}_{c}}$
unless otherwise stated, for simplicity.

We also define
\begin{eqnarray}
\chi &\equiv& \sqrt{\kappa}\varphi,\\
u & \equiv & \ln(e^\chi -1) \\
\Omega_{V_{0}} &\equiv& \frac{\kappa V_{0}}{3H_{0}^{2}},
\end{eqnarray}
to be used below.

Making discretized version of the above equations for $N$-body
simulations is non-trivial task. For example, the use of variable
$u$ instead of $\varphi$ (Appendix~\ref{appen:discret}) helps to
prevent $\varphi<0$, which is unphysical, but numerically possible
due to discretization. We refer the interested readers to
Appendix~\ref{appen:discret} to the whole treatment, with which we
can now proceed to do $N$-body runs.

\label{subsect:eqn_nonrel}

\section{Simulation and Results}

\label{sect:simu}

\subsection{Simulation Details}

\label{subsect:simu_detail}

\subsubsection{The $N$-body Code: MLAPM}

The full name of MLAPM is Multi-Level Adaptive Particle Mesh code.
As the name has suggested, this code uses multilevel grids
\cite{Brandt1977, Press1992, Briggs2000} to accelerate the
convergence of the (nonlinear) Gauss-Seidel relaxation method
\cite{Press1992} in solving boundary value partial differential
equations. But more than this, the code is also adaptive, always
refining the grid in regions where the mass/particle density
exceeds a certain threshold. Each refinement level form a finer
grid which the particles will be then (re)linked onto and where
the field equations will be solved (with a smaller time step).
Thus MLAPM has two kinds of grids: the domain grid which is fixed
at the beginning of a simulation, and refined grids which are
generated according to the particle distribution and which are
destroyed after a complete time step.

One benefit of such a setup is that in low density regions where
the resolution requirement is not high, less time steps are
needed, while the majority of computing sources could be used in
those few high density regions where high resolution is needed to
ensure precision.

Some technical issues must be taken care of however. For example,
once a refined grid is created, the particles in that region will
be linked onto it and densities on it are calculated, then the
coarse-grid values of the gravitational potential are interpolated
to obtain the corresponding values on the finer grid. When the
Gauss-Seidel iteration is performed on refined grids, the
gravitational potential on the boundary nodes are kept constant
and only those on the interior nodes are updated according to
Eq.~(\ref{eq:GS}): just to ensure consistency between coarse and
refined grids. This point is also important in the scalar field
simulation because, like the gravitational potential, the scalar
field value is also evaluated on and communicated between
multi-grids (note in particular that different boundary conditions
lead to different solutions to the scalar field equation of
motion).

In our simulation the domain grid (the finest grid that is not a
refined grid) has $128^{3}$ nodes, and there are a ladder of
coarser grids with $64^3$, $32^3$, $16^3$, $8^3$, $4^3$ nodes
respectively. These grids are used for the multi-grid acceleration
of convergence: for the Gauss-Seidel relaxation method, the
convergence rate is high upon the first several iterations, but
quickly becomes very slow then; this is because the convergence is
only efficient for the high frequency (short-range) Fourier modes,
while for low frequency (long-range) modes more iterations just do
not help much. To accelerate the solution process, one then
switches to the next coarser grid for which the low frequency
modes of the finer grid are actually high frequency ones and thus
converge fast. The MLAPM solver adopts the self-adaptive scheme:
if convergence is achieved on a grid, then interpolate the
relevant quantities back to the finer grid (provided that the
latter is not on the refinements) and solve the equation there
again; if convergence becomes slow on a grid, then go to the next
coarser grid. This way it goes indefinitely except when converged
solution on the domain grid is obtained or when one arrives at the
coarsest grid (normally with $2^3$ nodes) on which the equations
can be solved exactly using other techniques. For our scalar field
model, the equations are difficult to solve anyway, and so we
truncate the coarser-grid series at the $4^3$-node one, on which
we simply iterate until convergence is achieved. Furthermore, we
find that with the self-adaptive scheme in certain regimes the
nonlinear GS solver tends to fall into oscillations between
coarser and finer grids; to avoid such situations, we then use
V-cycle \cite{Press1992} instead.

For the refined grids the method is different: here one just
iterate Eq.~(\ref{eq:GS}) until convergence, without resorting to
coarser grids for acceleration.

As is normal in the Gauss-Seidel relaxation method, convergence is
deemed to be achieved when the numerical solution $u^{k}_{n}$
after $n$ iterations on grid $k$ satisfies that the norm
$\Vert\cdot\Vert$ (mean or maximum value on a grid) of the
residual
\begin{eqnarray}
e^{k} &=& L^{k}(u^{k}_{n}) - f_{k},
\end{eqnarray}
is smaller than the norm of the truncation error
\begin{eqnarray}
\tau^{k} &=& L^{k-1}(\mathcal{R}u^{k}_{n}) -
\mathcal{R}\left[L^{k}(u^{k}_{n})\right]
\end{eqnarray}
by a certain amount, or, in the V-cycle case, the reduction of
residual after a full cycle becomes smaller than a predefined
threshold (indeed the former is satisfied whenever the latter is).
Note here $L^{k}$ is the discretization of the differential
operator Eq.~(\ref{eq:diffop}) on grid $k$ and $L^{k-1}$ a similar
discretization on grid $k-1$, $f_k$ is the source term,
$\mathcal{R}$ is the restriction operator to interpolate values
from the grid $k$ to the grid $k-1$. In the modified code we have
used the full-weighting restriction for $\mathcal{R}$.
Correspondingly there is a prolongation operator $\mathcal{P}$ to
obtain values from grid $k-1$ to grid $k$, and we use a bilinear
interpolation for it. For more details see \cite{MLAPM}.

MLAPM calculates the gravitational forces on particles by centered
difference of the potential $\Phi$ and propagate the forces to
locations of particles by the so-called triangular-shaped-cloud
(TSC) scheme to ensure momentum conservation on all grids. The TSC
scheme is also used in the density assignment given the particle
distribution.

The main modifications to the MLAPM code for our model are:
\begin{enumerate}
    \item We have added a parallel solver for the scalar field
    based on Eq.~(\ref{eq:u_phi_EOM}). The solver uses a
    nonlinear Gauss-Seidel method and the same criterion for
    convergence as the (linear) Gauss-Seidel Poisson solver.
    \item The solved value of $u$ is then used to calculate local
    mass density and thus the source term for the Poisson
    equation, which is solved using fast Fourier transform.
    \item The fifth force is obtained by differentiating the $u$
    just like the calculation of gravity.
    \item The momenta and positions of particles are then updated
    taking in account of both gravity and the fifth force.
\end{enumerate}
There are a lot of additions and modifications to ensure smooth
interface and the newly added data structures. For the output, as
there are multilevel grids all of which host particles, the
composite grid is inhomogeneous and thus we choose to output the
positions, momenta of the particles, plus the gravity, fifth force
and scalar field value \emph{at the positions} of these particles.
We can of course easily read these data into the code, calculate
the corresponding quantities on each grid and output them if
needed.

\subsubsection{Inclusion of Baryons}

As mentioned above, the most important difference of the present
work from \cite{Li2009} is the inclusion of baryons - the
particles which do not couple to the scalar field. The baryons do
not contribute to the scalar field equation of motion and are not
affected by the scalar fifth force, at least directly, so that it
is important to make sure that they do not mess up the physics.

In the modified code we distinguish baryons and CDM particles by
tagging all of them. We consider the situation where 20\% of all
matter particles are baryonic and 80\% are CDM. At the beginning
of each simulation, we loop over all particles and for each
particle we generate a random number from a uniform distribution
in $[0,1]$. If this random number is less than 0.2 then we tag the
particle as baryon, and otherwise we tag it as CDM. Once these
tags have been set up they will never been changed again, and the
code then determines whether the particle contributes to the
scalar field evolution and feels the fifth force or not according
to its tag.

\subsubsection{Initial Condition and Simulation Parameters}

All the simulations are started at the redshift $z=49$. In
principle, modified initial conditions (initial displacements and
velocities of particles which is obtained given a linear matter
power spectrum) need to be generated for the coupled scalar field
model, because the Zel'dovich approximation \cite{Zeldovich,
Efstathiou1985} is also affected by the scalar field coupling
\cite{Baldi2008}. In practice, however, we have found in our
linear perturbation calculation \cite{Li2009} that the effect on
the linear matter power spectrum is negligible
($\lesssim\mathcal{O}(10^{-4})$) for our choices of parameters
$\gamma, \mu$. Another way to see that the scalar field has really
negligible effects on the matter power spectrum at early times is
to look at Fig.~\ref{fig:Figure2} below, which shows that at those
times the fifth force is just much weaker than gravity and
therefore its impact ignorable. Considering these, we simply use
the $\Lambda$CDM initial displacements/velocities for the
particles in these simulations, which are generated using GRAFIC2
\cite{GRAFIC}.

\begin{figure*}[tbp]
\centering \includegraphics[scale=0.7] {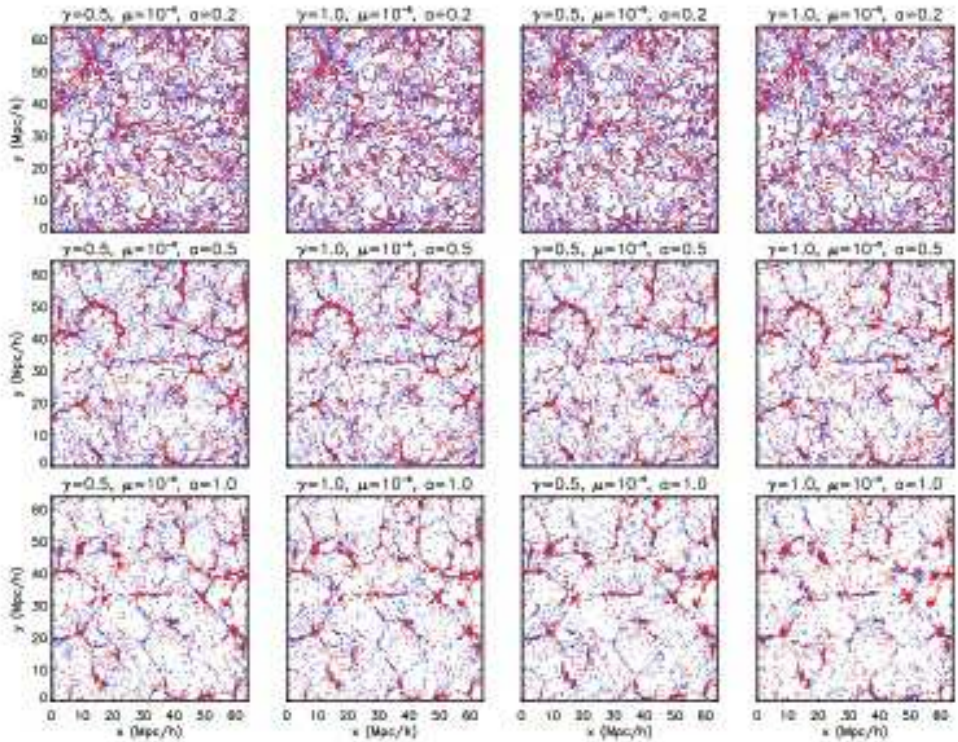}
\caption{(Colour Online) The snapshots of the distribution of
particles. The 4 columns are for the 4 models we consider and the
3 rows are for three output times, respectively $a=0.2, 0.5$ and
$1$ (corresponding to redshifts 4, 1 and 0) where $a$ is the
cosmic scale factor. The red crosses denote baryons while blue
dots represent CDM particles. All particles are take from a slice
of the simulation box with $32.0h^{-1}~\mathrm{Mpc} < z <
32.1h^{-1}~\mathrm{Mpc}$ and projected to an $x-y$ plane.}
\label{fig:Figure1}
\end{figure*}

The physical parameters we use in the simulations are as follows:
the present-day dark energy fractional energy density
$\Omega_{\mathrm{DE}}=0.76$ and
$\Omega_{m}=\Omega_{\mathrm{CDM}}+\Omega_{\mathrm{B}}=0.24$,
$H_{0}=73$~km/s/Mpc, $n_{s}=0.958$, $\sigma_{8}=0.8$. The
simulation box has a size of $64h^{-1}$~Mpc, where
$h=H_0/(100~\mathrm{km/s/Mpc})$. We simulate 4 models, with
parameters $(\gamma,\mu)$ equal to $(0.5,10^{-6})$,
$(0.5,10^{-5})$, $(1.0,10^{-6})$ and $(1.0,10^{-5})$ respectively
(such parameters are chosen so that the deviation from
$\Lambda$CDM will be neither too small to be distinguishable or
too large to be realistic). For each model we make 5 runs with
exactly the same initial condition, but different seeds in
generating the random number to tag baryons and CDM particles; all
the 4 models use the same 5 seeds so that results can be directly
compared. We hope the average of the results from 5 runs could
reduce the scatter. In all those simulations the mass resolution
is $1.04\times10^{9}h^{-1}~M_{\bigodot}$, the particle number is
$256^{3}$, the domain grid is a $128\times128\times128$ cubic and
the finest refined grids have 16384 cells on one side,
corresponding to a force resolution of $\sim 12h^{-1}~$kpc.

We also make a run for the $\Lambda$CDM model using the same
parameters (except for $\mu, \gamma$, which are not needed now)
and initial condition.

\subsection{Preliminary Results}

\label{subsect:simu_result}

In Table~\ref{tab:table1} we have listed some of the main results
for the 20 runs we have made, from which we could obtain some
rough idea how the motions of baryons and CDM particles differ
from each other. We see that for the model $\gamma=1.0,
\mu=10^{-5}$ the CDM particles could be up to $\sim1.6$ times
faster than baryons, thanks to the enhancement by the fifth force.
We will come back to this point later when we argue for the
necessity of a modified strategy of identifying virialized halos.
\begin{table}[htbp]
\caption{\label{tab:table1} The mean velocity of all particles
($\bar{v}$), baryons ($\bar{v}_{\mathrm{B}}$) and CDM particles
($\bar{v}_{\mathrm{CDM}}$) at $z=0$ for the 20 coupled scalar
field runs. The unit is proper km/s.} \label{table}
\end{table}
\begin{center}
\begin{tabular}{ccccc}
\hline\hline
parameters & run no. & $\bar{v}$ & $\bar{v}_{B}$ & $\bar{v}_{\mathrm{CDM}}$ \\
\hline
 & 1 & 367.68\ \ & 337.71\ \ & 375.18 \\
 & 2 & 367.70\ \ & 337.69\ \ & 375.19 \\
$\gamma=0.5, \mu=10^{-6}$ & 3 & 367.75\ \ & 337.63\ \ & 375.28 \\
 & 4 & 367.67\ \ & 337.76\ \ & 375.15 \\
 & 5 & 367.65\ \ & 337.59\ \ & 375.16 \\
\hline
 & 1 & 382.12\ \ & 335.91\ \ & 393.68 \\
 & 2 & 382.01\ \ & 335.79\ \ & 393.55 \\
$\gamma=1.0, \mu=10^{-6}$ & 3 & 381.91\ \ & 335.79\ \ & 393.55 \\
 & 4 & 382.01\ \ & 335.87\ \ & 393.54 \\
 & 5 & 382.05\ \ & 335.83\ \ & 393.61 \\
\hline
 & 1 & 412.75\ \ & 357.36\ \ & 426.60 \\
 & 2 & 412.75\ \ & 357.49\ \ & 426.55 \\
$\gamma=0.5, \mu=10^{-5}$ & 3 & 426.60\ \ & 357.44\ \ & 438.03 \\
 & 4 & 412.92\ \ & 357.62\ \ & 426.75 \\
 & 5 & 412.73\ \ & 357.44\ \ & 426.56 \\
\hline
 & 1 & 565.22\ \ & 388.77\ \ & 609.35 \\
 & 2 & 565.13\ \ & 388.68\ \ & 609.19 \\
$\gamma=1.0, \mu=10^{-5}$ & 3 & 565.19\ \ & 388.82\ \ & 609.30 \\
 & 4 & 565.58\ \ & 388.25\ \ & 609.66 \\
 & 5 & 564.64\ \ & 388.75\ \ & 608.63 \\
\hline\hline
\end{tabular}
\end{center}

\begin{figure*}[tbp]
\centering \includegraphics[scale=0.7] {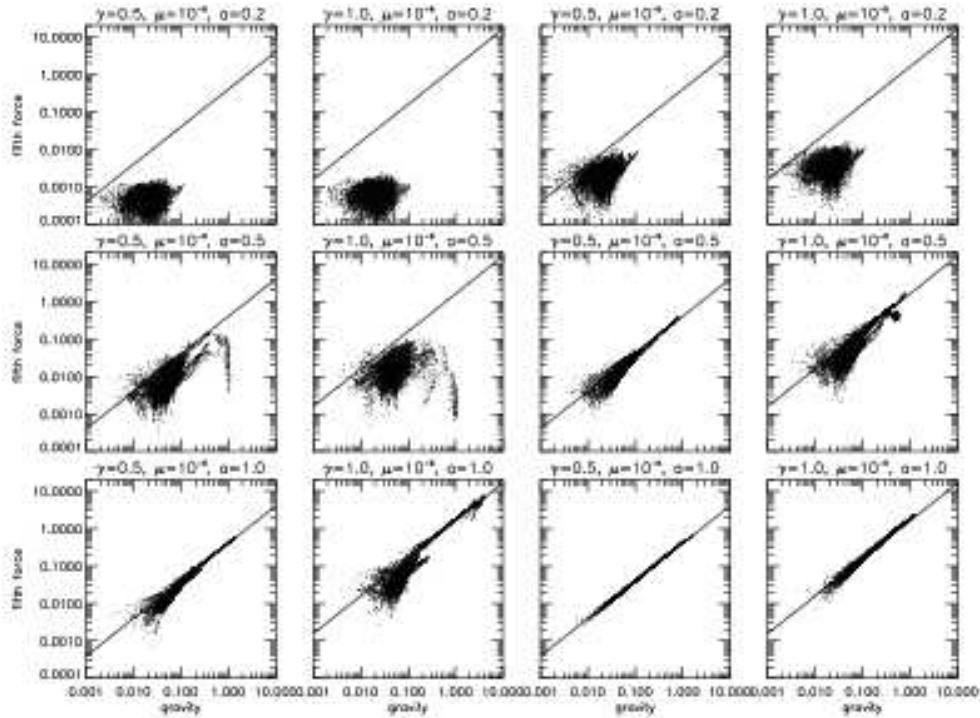}
\caption{The correlation of the magnitudes of the fifth force and
gravity on the \emph{CDM} particles. The dots denote CDM particles
taken from a slice of simulation box with $32.0h^{-1}~\mathrm{Mpc}
< z < 32.1h^{-1}~\mathrm{Mpc}$, and the lines are the predicted
relation between fifth force and gravity should the former be not
suppressed by the chameleon effect. The 4 columns are for the 4
models we consider and the 3 rows are for three output times,
respectively $a=0.2, 0.5$ and $1$ (corresponding to redshifts 4, 1
and 0) where $a$ is the cosmic scale factor.} \label{fig:Figure2}
\end{figure*}

\begin{figure*}[tbp]
\centering \includegraphics[scale=0.7] {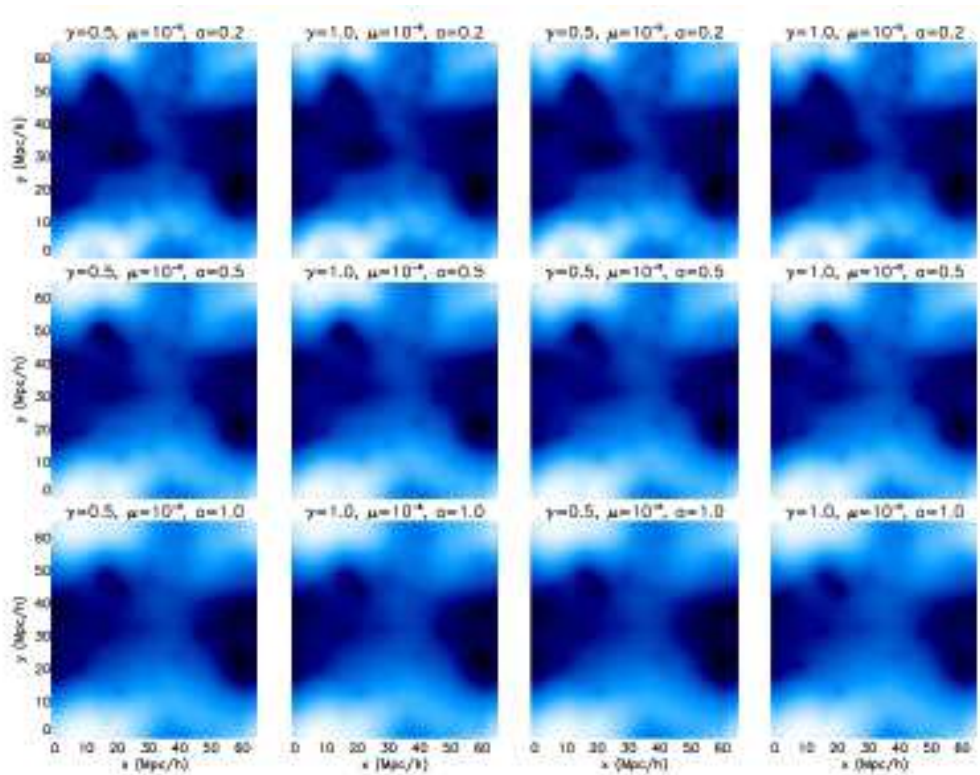}
\caption{(Colour Online) The gravitational potential on the $z$
plane where $z=32h^{-1}$~Mpc. Dark regions are where the potential
is more negative (deeper) while light regions are where it is less
negative (shallower). The 4 columns are for the 4 models we
consider and the 3 rows are for three output times, respectively
$a=0.2, 0.5$ and $1$ (corresponding to redshifts 4, 1 and 0) where
$a$ is the cosmic scale factor.} \label{fig:Figure3}
\end{figure*}

\begin{figure*}[tbp]
\centering \includegraphics[scale=0.7] {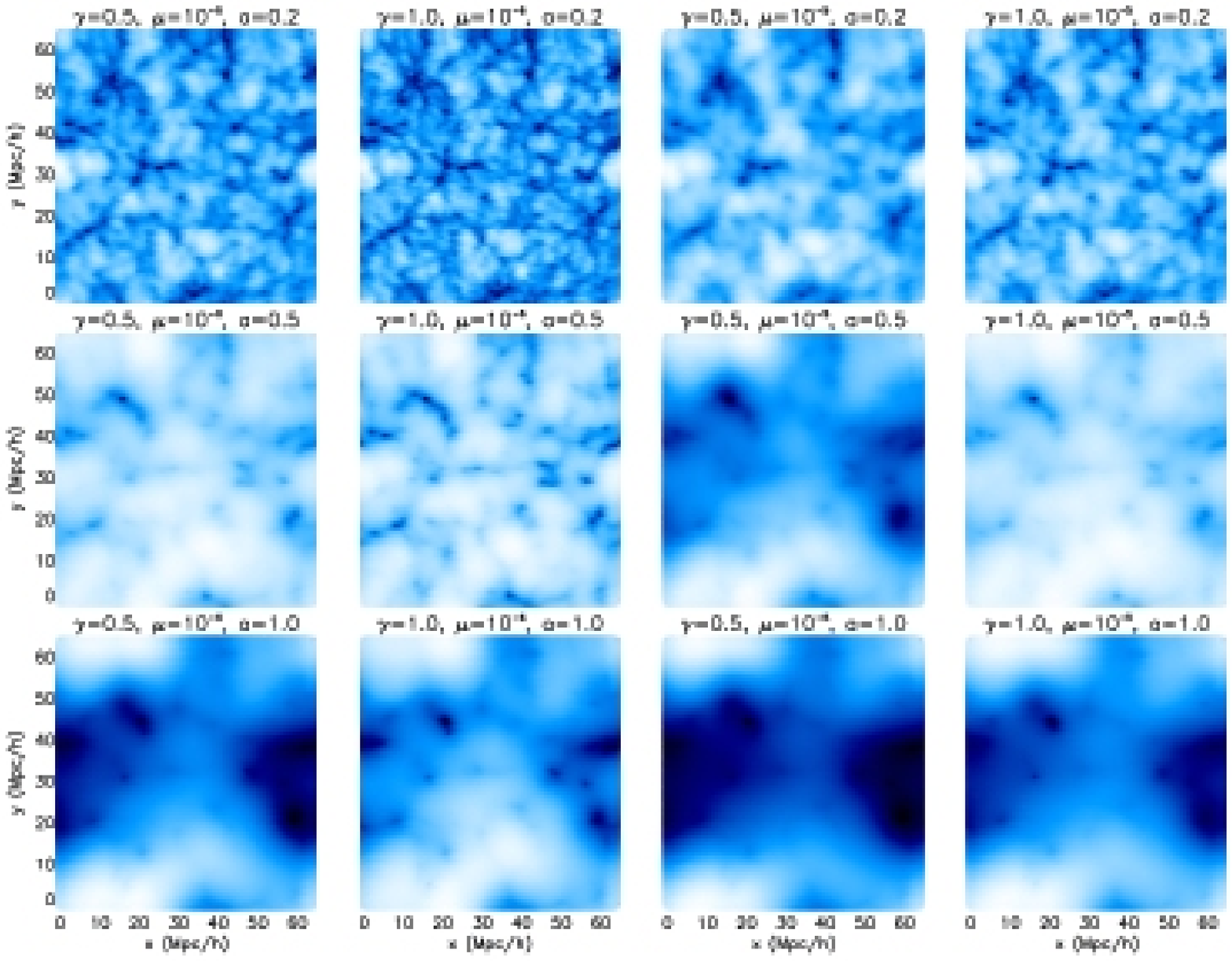}
\caption{(Colour Online) The value of the scalar field $\varphi$
on the $z$ plane where $z=32h^{-1}$~Mpc. Plotted is
$\log(\sqrt{\kappa}\varphi)$ for better contrast. Dark regions are
where $\varphi$ is closer to 0 while light regions are where it is
closer to the background value. The 4 columns are for the 4 models
we consider and the 3 rows are for three output times,
respectively $a=0.2, 0.5$ and $1$ (corresponding to redshifts 4, 1
and 0) where $a$ is the cosmic scale factor.} \label{fig:Figure4}
\end{figure*}

In Fig.~\ref{fig:Figure1} we have shown some snapshots of the
distribution of baryonic and CDM particles, to give some idea
about the hierachical structure formation and for comparisons with
other figures below. It shows clearly how some clustering objects
develop, with filaments connecting them together. The baryons
roughly follow the clustering of CDM particles, but in some low
density regions they become slightly separated.

To understand how the motion of the particles is altered by the
coupling to the scalar field, in Fig.~\ref{fig:Figure2} we have
shown the correlation between the magnitudes of the fifth force
and gravity on the CDM particles (remember that baryons do not
feel the fifth force). Ref.~\cite{Li2009} has made a detailed
qualitatively analysis about the general trend of this correction,
and here we just give a brief description:

From Eqs.~(\ref{eq:INTPoisson}, \ref{eq:INTphiEOM}) we could see
that, \emph{when the scalar field potential}, \emph{i.e.},
\emph{the last term of Eqs.~(\ref{eq:INTPoisson},
\ref{eq:INTphiEOM}) could be neglected}, then the scalar field
$\varphi$ is simply proportional to the gravitational potential
$\Phi$ and as a result Eq.~(\ref{eq:INTdpdtcomov}) tells us that
the strength of the fifth force is just $2\gamma^{2}$ times that
of gravity; in other words, the effect of the scalar field is a
rescaling of the gravitational constant by $1+2\gamma^{2}$. This
is because in this situation, the effective mass of the scalar
field, which is given by
$m^{2}_{eff}=\partial^{2}V_{eff}/\partial\varphi^{2}$, where
\begin{eqnarray}
V_{eff}(\varphi) &=& V(\varphi)+C(\varphi)\rho_{\mathrm{CDM}}
\end{eqnarray}
is the effective total potential, is light and the fifth force is
long-range, like gravity. For comparison, in
Fig.~\ref{fig:Figure2} we also plot this $2\gamma^{2}$ proportion
between the two forces, as a straight line:
$\lg(f)=\lg(0.8g)+\lg(2\gamma^{2})$, where $f, g$ denote
respectively the magnitudes of the fifth force and gravity, and
the factor 0.8 in front of $g$ comes from the fact that only 80\%
of the particles are CDM (and thus contribute to the fifth force).
This scaling relation actually sets an upper limit on how strong
the fifth force could be relative to gravity, should it not be
suppressed by other effects.

In contrast, when the value of $\varphi$ is small, the last term
of Eqs.~(\ref{eq:INTPoisson}, \ref{eq:INTphiEOM}) is not
negligible and the scalar field acquires a heavy mass, making it
short-ranged. As a result, a particle outside a high density
region might not feel the fifth force exerted by particles in that
region, even it is quite close to the region. But because it can
feel gravity from that region, so the total fifth force on the
particle becomes less than the $2\gamma^{2}$ scaling.

In general, the value of $\varphi$ is determined by $\mu, \gamma,
\rho_{\mathrm{CDM}}$ as well as its background value
$\bar{\varphi}$ (which sets the boundary condition to solve the
interior value). At early times $\bar{\varphi}$ is very close to 0
and $\rho_{\mathrm{CDM}}$ is high everywhere, making $\varphi$
small everywhere too, and suppressing the fifth force so that it
is significantly below the $2\gamma^{2}$-scaling (first row of
Fig.~\ref{fig:Figure2}). At later times, $\bar{\varphi}$ increases
and $\rho_{\mathrm{CDM}}$ decreases, weakening the above effect so
that the fifth force becomes "saturated" (\emph{i.e.}, approaches
the $2\gamma^2$ prediction) and the points in the figure hit the
straight lines (last two rows). Because decreasing $\mu$ and
increasing $\gamma$ have the same effects of making $\varphi$
small, in the models with $\mu=10^{-6}$ the fifth force saturates
later than in the models with $\mu=10^{-5}$. In addition, because
high $\rho_{\mathrm{CDM}}$ tends to decrease $\varphi$ and
increase the scalar field mass, so in high density regions (where
gravity is stronger) the fifth force also saturates later.

The agreement between the numerical solution of the fifth force
and the $2\gamma^{2}$-scaling relation in cases of weak-chameleon
effect serves as an independent check of our numerical code.

Fig.~\ref{fig:Figure3} plots the spatial configuration for the
gravitational potential at the same position and output times as
in Fig.~\ref{fig:Figure1}. As expected, the potential is
significantly deeper where there is significant clustering of
matter [cf.~Fig.~\ref{fig:Figure1}].

We also show in Fig.~\ref{fig:Figure4} the spatial configuration
for the scalar field $\varphi$ at the same output position and
times. At early times when $\varphi$ is small and the scalar field
mass is heavy, the fifth force is so short-ranged that $\varphi$
only depends on the local density. This means that the spatial
configuration of $\varphi$ in this situation could well reflect
the underlying dark matter distribution, a fact which could be
seen clearly in the first row. As time passes by, the mass of the
scalar field decreases on average and $\bar{\varphi}$ increases,
the value of $\varphi$ at one point is more and more influenced by
the matter distribution in neighboring regions, and such an
"averaging" effect weakens the contrast and makes the plots
blurring (last two rows). Furthermore, obviously decreasing $\mu$
and increasing $\gamma$ could increase the scalar field's mass,
shorten the range of the fifth force, make $\varphi$ less
dependent on its value in neighboring regions, and thus strengthen
the contrast in the figures.

\subsection{Power Spectrum}

\label{subsect:simu_ps}

To have a more quantitative description about how the matter
clustering property is modified with respect to the $\Lambda$CDM
model, we consider the matter power spectra $P(k)$ in our
simulation boxes.

The nonlinear matter power spectrum in the present work is
measured using POWMES \cite{powmes}, which is a public available
code based on the Taylor expansion of trigonometric functions and
yields Fourier modes from a number of fast Fourier transforms
controlled by the order of the expansion. We also average the
results from the 5 runs for each model and calculate the variance.

\begin{figure*}[tbp]
\centering \includegraphics[scale=1.05] {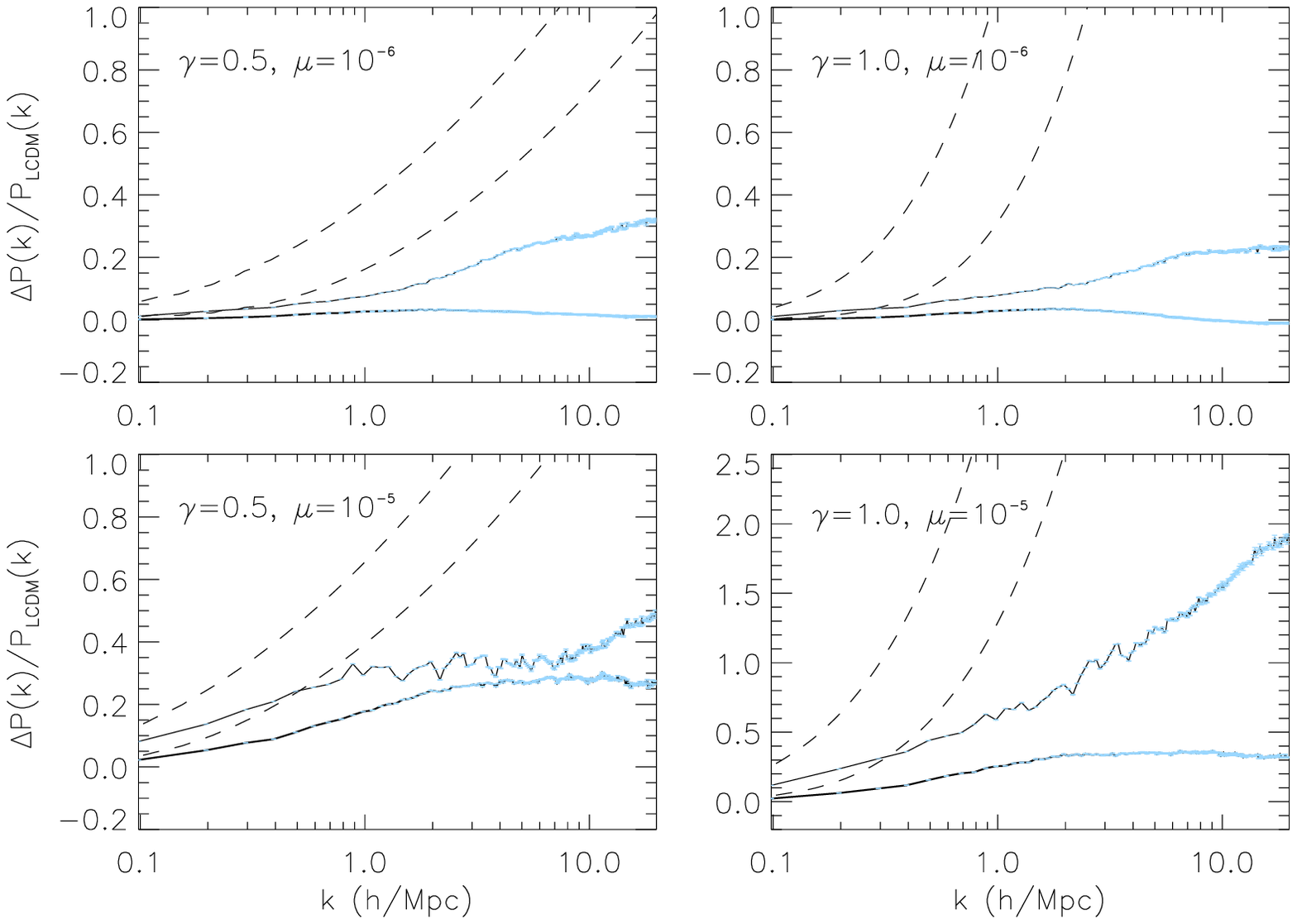}
\caption{The fractional difference between the nonlinear matter
power spectra for the inhomogeneously coupled scalar field model
and the $\Lambda$CDM paradigm, where $\Delta P(k)$ is their
difference. The range of $k$ is 0.1 to 50 $h$~Mpc$^{-1}$. The
solid curves are mean and the light-blue error bars the standard
deviation calculated from the 5 samples of measurements. Upper
solid curves in each panel: the result when $a=1.0$ (redshift 0);
Lower solid curves in each panel: the result for $a=0.5$ (redshift
1). The dashed curves are the corresponding results from the
linear perturbation theory for comparison.} \label{fig:Figure5}
\end{figure*}

\begin{figure*}[tbp]
\centering \includegraphics[scale=1.05] {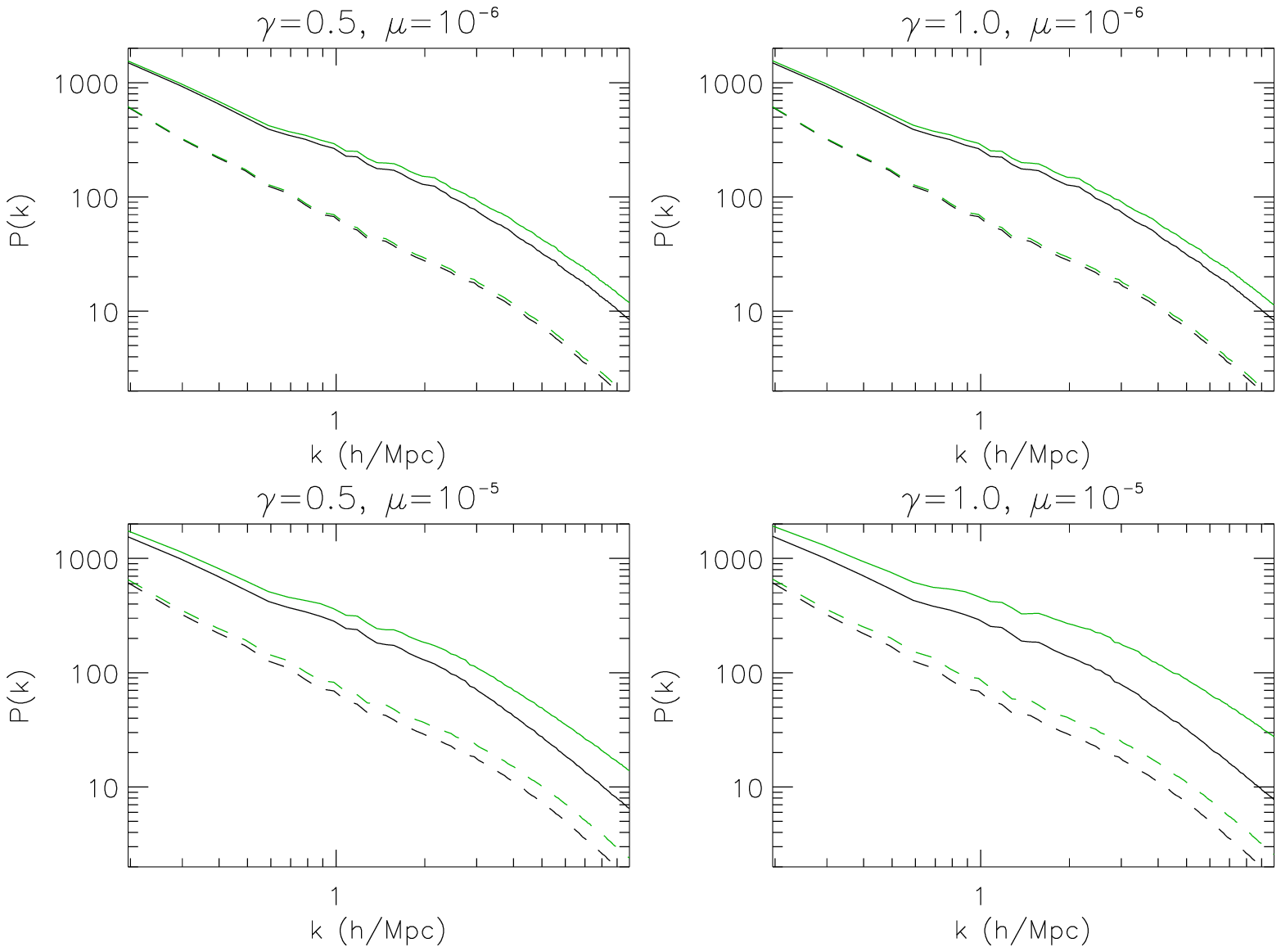}
\caption{The nonlinear power spectra for CDM and baryonic
particles. Shown are the results at two different output times:
$a=0.5$ (dashed curves) and $a=1.0$ (solid curves). The upper
curves are for CDM particles the lower curves for baryons. Note
that we only plot the results in the range
$0.1~h~\mathrm{Mpc}^{-1} \leq k \leq 10~h~\mathrm{Mpc}^{-1}$ due
to the small number of baryonic particles. Note also that all the
curves are the mean of 5 relevant runs, but we have not included
the standard deviations as in other plots, because they are small
and only make the curves unrecognizable.} \label{fig:Figure6}
\end{figure*}

In Fig.~\ref{fig:Figure5} shown are the fractional differences
between the $P(k)$ for our 4 models and for $\Lambda$CDM, at two
different output times. At early times ($a=0.5$, upper solid
curves in each panel), the difference is generally small, but
still the 2 models with $\mu=10^{-5}$ show up to $\sim25\%$
deviation from $\Lambda$CDM prediction. This is because for larger
$\mu$ the scalar field is lighter and the fifth force less
suppressed, its influence in the structure formation therefore
enhanced. Notice that on small scales the deviation from
$\Lambda$CDM decreases, which is a desirable property of chameleon
models which are designed to suppress the fifth force on small
scale high density regions.

The lower solid curves in each panel of Fig.~\ref{fig:Figure5}
display the same quantities at $a=1.0$ (late times). We can see
the trend of increasing deviation from $\Lambda$CDM for all 4
models, because fifth force is essentially unsuppressed at the
late epoch [cf.~Fig.~\ref{fig:Figure2}]. For example, the
deviation of the model $\gamma=1.0, \mu=10^{-5}$ is significantly
larger than that of the model $\gamma=0.5, \mu=10^{-5}$ as
na\"{\i}vely expected, thanks to the lack of suppression of fifth
force in both models (the $\gamma=0.5$ model obviously has a
smaller saturated fifth force).

For comparison we also plot the $\Delta P/P$ that is predicted by
the linear perturbation theory for the 4 models under
consideration (the dashed curves). As can be seen there, at large
scales (small $k$) where linear perturbation is considered as a
good approximation, the linear and nonlinear results agree pretty
well (especially for the $a=0.5$ case). The largest scale we can
probe is limited by the size of our simulation boxes ($64~Mpc/h$)
and as a result we cannot make plot beyond the point $k\sim
0.1~h/Mpc$, where nonlinearity is expected to first become
significant. However, in the case of $a=1.0$, we can see the clear
trend of the linear and nonlinear results merging towards
$k<0.1~h/Mpc$ at vanishing $\Delta P/P$. Similar results can be
found in Fig.~2 of \cite{Oyaizu2008b} for f(R) gravity.

Because we have two species of matter particles, one uncoupled to
the scalar field, we are also interested in their respect power
spectrum and the bias between them. These are displayed in
Fig.~\ref{fig:Figure6}. The results could be understood easily:
because a CDM particle always feel stronger total force than a
baryon at the same position, so the clustering of the former is
identically stronger than the latter as well. This could result in
a significant bias between these two species at the present time,
especially for the models with $\mu=10^{-5}$, where the fifth
force is less suppressed.

\subsection{Mass Functions}

\label{subsect:simu_mf}

We identify halos in our $N$-body simulations using MHF (MLAPM
Halo Finder) \cite{mhf}, which is the default halo finder for
MLAPM. MHF optimally utilizes the refinement structure of the
simulation grids to pin down the regions where potential halos
reside and organize the refinement hierarchy into a tree
structure. Because MLAPM refines grids according to the particle
density on them, so the boundaries of the refinements are simply
isodensity contours. MHF collect the particles within these
isodensity contours (as well as some particles outside). It then
performs the following operations: (i) assuming spherical symmetry
of the halo, calculate the escape velocity $v_{esc}$ at the
position of each particle, (ii) if the velocity of the particle
exceeds $v_{esc}$ then it does not belong to the virialized halo
and is removed. (i) and (ii) are then iterated until all unbound
particles are removed from the halo or the number of particles in
the halo falls below a pre-defined threshold, which is 20 in our
simulations. Note that the removal of unbound particles is not
used in some halo finders using the spherical overdensity (SO)
algorithm, which includes the particles in the halo as long as
they are within the radius of a virial density contrast. Another
advantage of MHF is that it does not require a pre-defined linking
length in finding halos, such as the friend-of-friend procedure.

\begin{figure*}[tbp]
\centering \includegraphics[scale=0.8] {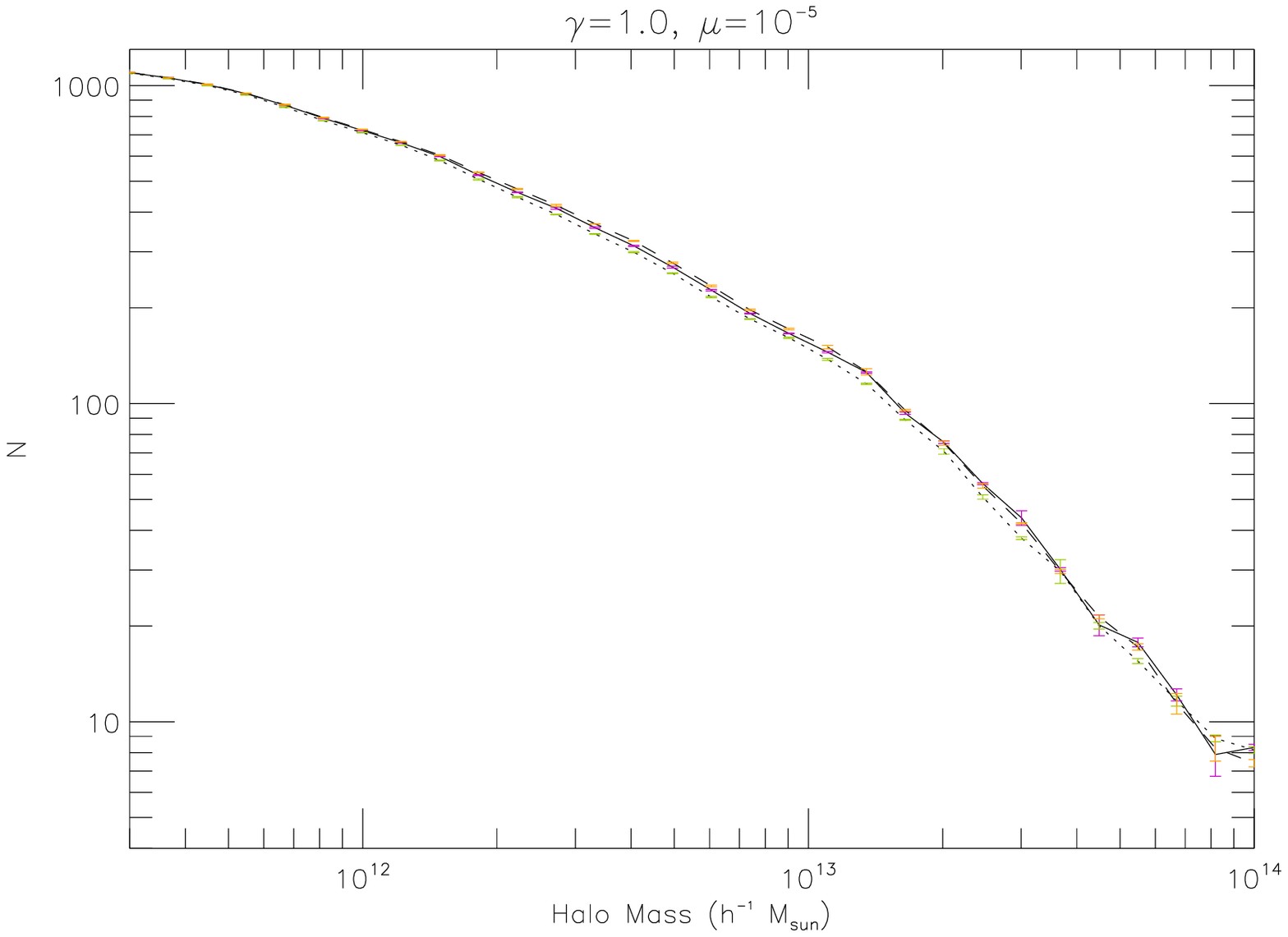}
\caption{The comparison of three different methods to identify
halos from the simulations. The first method (solid curve) uses
the approximation Eq.~(\ref{eq:ahf_vescnew}) to estimate the
escape velocity; the second method (dotted curve) uses the
$\Lambda$CDM prediction, which underestimates $v_{esc}$ and causes
more particles to be removed; the third method (dashed curve) uses
$v_{esc}=\infty$ so that no particles are removed. All curves are
for the model $\gamma=1.0, \mu=10^{-5}$, and the curves and error
bars denote respectively the mean and standard deviation for 5
sample results.} \label{fig:Figure7}
\end{figure*}

\begin{figure*}[tbp]
\centering \includegraphics[scale=1.05] {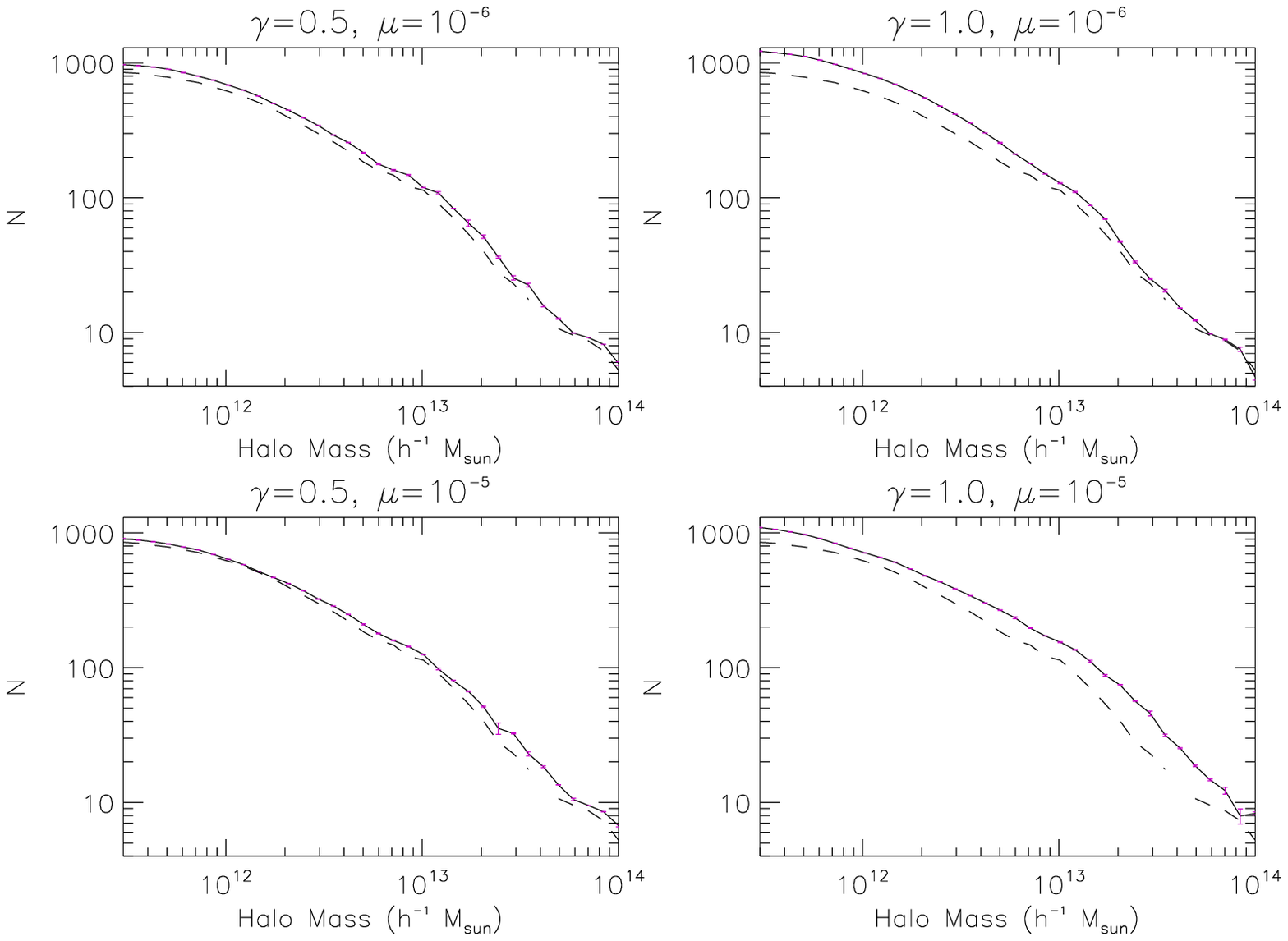}
\caption{The mass functions for the 4 models under consideration
at $a=1.0$ (solid curves). The dashed curve is the mass function
for the $\Lambda$CDM simulation. The (solid) curves and error bars
denote respectively the mean and standard deviation for 5 sample
results.} \label{fig:Figure8}
\end{figure*}

When it comes to our coupled scalar field model, the MHF algorithm
also needs to be modified. The reason is that, as we mentioned
above, the scalar field $\varphi$ behaves as an extra "potential"
(which produces the fifth force) and so the CDM particles
experience a deeper total "gravitational" potential than what they
do in the $\Lambda$CDM model. Consequently, the escape velocity
for CDM particles increases compared with the Newtonian
prediction. This is indeed important to bear in mind because, as
we have seen in Table~\ref{tab:table1}, the CDM particles are
typically much faster than what they are in the $\Lambda$CDM
simulation, and so if we underestimate $v_{esc}$ then some
particles which should have remained in the halo would be
incorrectly removed by MHF. In general, similar things should be
taken care of in other theories involving modifications to gravity
in the non-relativistic limit, such as MOND and $f(R)$ gravity.

An exact calculation of the escape velocity in the coupled scalar
field model is obviously difficult due to the complicated
behaviour of the scalar field, and thus here we introduce an
approximated algorithm, which is based on the MHF default method
\cite{ahf}, to estimate it.

MHF works out $v_{esc}$ using the Newtonian result $v^{2}_{esc} =
2|\Phi|$ in which $\Phi$ is the gravitational potential. Under the
assumption of spherical symmetry, the Poisson equation
$\nabla^{2}\Phi=4\pi G\rho_{m}$ could be integrated once to give
\begin{eqnarray}\label{eq:ahf_newton_law}
\frac{d\Phi}{dr} &=& \frac{GM(<r)}{r^{2}}
\end{eqnarray}
which is just the Newtonian force law. This equation can be
integrated once again to obtain
\begin{eqnarray}\label{eq:ahf_phi}
\Phi(r) &=& G\int^{r}_{0}\frac{M(<r')}{r'^{2}}dr' + \Phi_{0}
\end{eqnarray}
where $\Phi_{0}$ is an integration constant and can be fixed
\cite{ahf} by requiring that $\Phi(r=\infty)=0$ as
\begin{eqnarray}\label{eq:ahf_phi0}
\Phi_{0} &=& \frac{GM_{vir}}{R_{vir}} +
G\int^{R_{vir}}_{0}\frac{M(<r')}{r'^{2}}dr',
\end{eqnarray}
in which $R_{vir}$ is the virial radius of the halo and $M_{vir}$
is the mass enclosed in $R_{vir}$.

When the fifth force acts on CDM particles, the force law
Eq.~(\ref{eq:ahf_newton_law}) is modified and these particles feel
a larger total "gravitational" potential. To take this into
account, we need to have the knowledge about how the force law is
modified and a simple rescaling of gravitational constant $G$ has
been shown to be not physical in certain regimes.

To solve the problem, we notice that in the MHF code
Eq.~(\ref{eq:ahf_newton_law}) is used in the numerical
integrations to obtain both $\Phi(r)$ and $\Phi_{0}$
[cf.~Eqs.(\ref{eq:ahf_phi}, \ref{eq:ahf_phi0})]. More explicitly,
the code loops over all particles in the halo in ascending order
of the distance from the halo centre, and when a particle is
encountered its mass is uniformly distributed into the spherical
shell between the particle and its previous particle (the
thickness of the shell is then the $dr$ of the integration).
Obviously when the fifth force is added into
Eq.~(\ref{eq:ahf_newton_law}) we could use the same method to
compute the total "gravitational" potential which includes the
contribution from $\varphi$ (call this contribution
$\Phi_{\varphi}$; because $d\Phi_{\varphi}/dr =
\mathrm{fifth~force}$, so from Eq.~(\ref{eq:WFdpdtcomov}) we can
easily see $\Phi_{\varphi}=\frac{C_{,\varphi}}{C}a\varphi$). But
now there is a subtlety here: not all particles are CDM while only
CDM particles contribute to $\Phi_{\varphi}$. So in the modified
MHF code we calculate $\Phi, \Phi_{0}$ and $\Phi_{\varphi},
\Phi_{\varphi0}$ separately, using all particles for the former
and only CDM for the latter. Finally
\begin{eqnarray}\label{eq:ahf_vescnew}
v_{esc}^{2} &=& 2|\Phi+\Phi_{\varphi}-\Phi_{0}-\Phi_{\varphi0}|
\end{eqnarray}
is our estimate of the escape velocity. Because we have recorded
the components of gravity and the fifth force for each particle in
the simulation, so the fifth-force-to-gravity ratio can be
computed at the position of each particle, which is approximated
to be
$\left(\Phi_\varphi-\Phi_{\varphi0}\right)/\left(\Phi-\Phi_0\right)$
at that position. In this way we have, at least approximately,
taken into account the environment dependence of the fifth force
and thus of $\Phi_{\varphi}$.

To see how our modification of the MHF code affects the final
result on the mass function, in Fig.~\ref{fig:Figure7} we compare
the mass functions for the model $\gamma=1.0, \mu=10^{-5}$
calculated using three methods: the modified MHF, the original MHF
(by default for $\Lambda$CDM simulations) and another modified MHF
in which we set $v_{esc}$ to be very large so that no particles
will ever escape from any halo (this is similar to the spherical
overdensity algorithm mentioned above). It could be seen that the
difference between these methods can be up to a few percent for
small halos where the potential is shallow, particle number is
small and the result is sensitive to whether more particles are
removed. As expected, the second method gives least halos because
there $v_{esc}$ is smallest and many particles are removed, while
the third method gives most halos because no particles are removed
at all.

Fig.~\ref{fig:Figure8} displays the mass functions for the 4
models as compared to the $\Lambda$CDM result. As expected, the
fifth force enhances the structure formation and thus produces
more halos in the simulation box.

\subsection{Halo Profiles}

\label{subsect:simu_prof}

As our simulations have reached a force resolution of
$\sim12h^{-1}$~kpc, while the typical size of the large halos in
the simulations is of order Mpc, we could ask what the internal
profiles of the halos look like and how they have been modified by
the coupling between CDM particles and the scalar field.

\begin{figure*}[tbp]
\centering \includegraphics[scale=1.05] {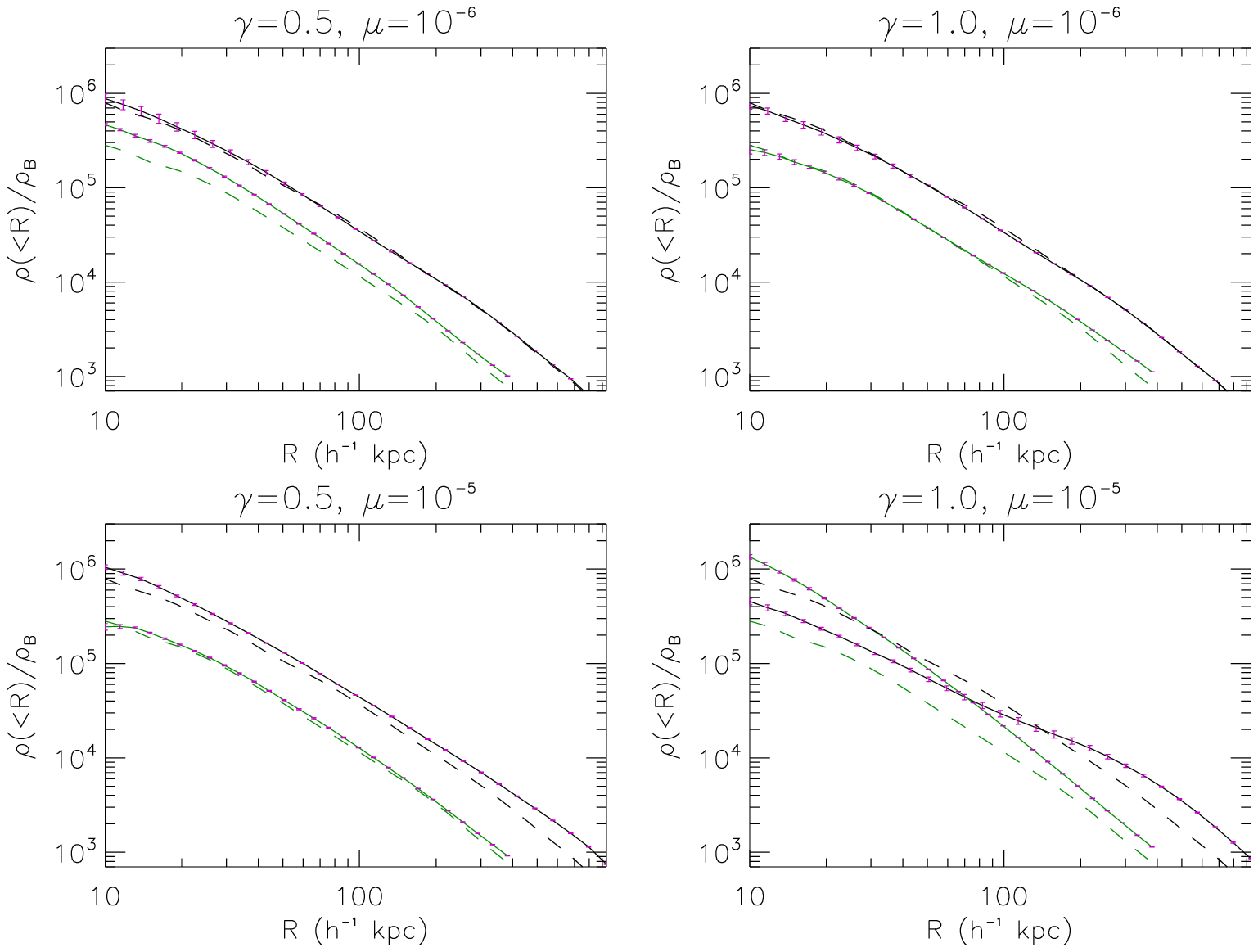}
\caption{The overdensity inside two typical halos selected from
the simulations for each model (see text for a more detailed
description), as a function of the halo radius $R$. The solid
curves and error bars denote respectively the mean and standard
deviation for the results from our 5 sample runs, while the dashed
curve is the $\Lambda$CDM prediction. In each panel the upper two
curves (black) at large radii are for halo I and the lower two
curves (green) at large radii are for halo II. $\rho_{\mathrm{B}}$
is the background matter (baryons and CDM) density. We skipped the
result for $R<10h^{-1}$~kpc, which is not reliable due to the
limit from resolution.} \label{fig:Figure9}
\end{figure*}

\begin{figure*}[tbp]
\centering \includegraphics[scale=0.7] {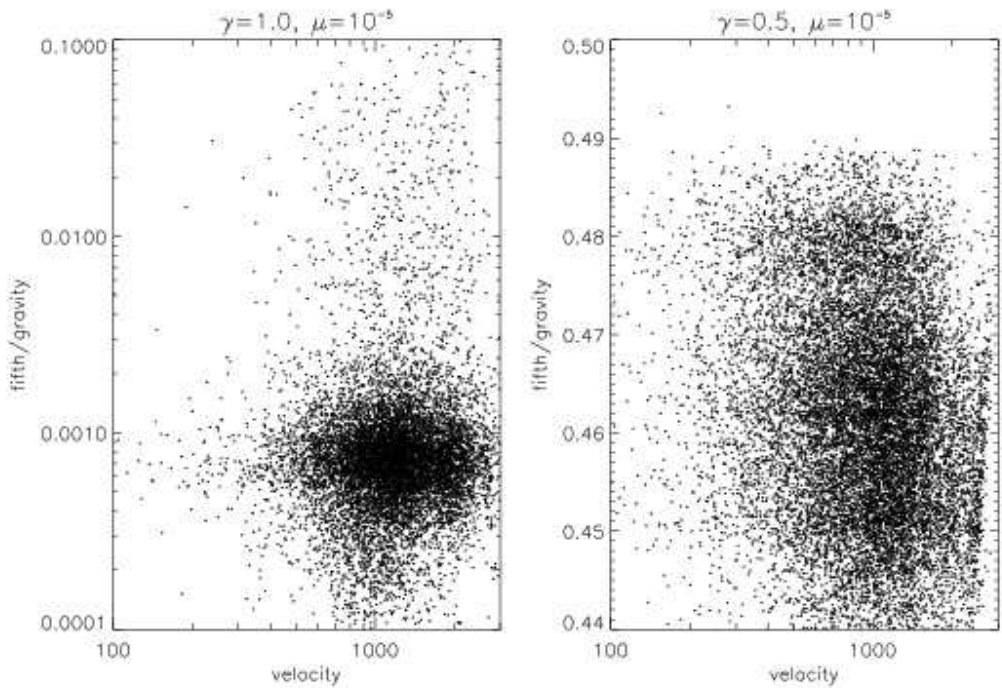}
\caption{Distribution of the velocities of the particles in halo I
and the fifth-force-to-gravity ratio at the positions of these
particles, for the two models $\gamma=0.5, \mu=10^{-5}$ and
$\gamma=1.0, \mu=10^{-5}$. Each dots represent a particle.}
\label{fig:Figure11}
\end{figure*}

\begin{figure*}[tbp]
\centering \includegraphics[scale=1.05] {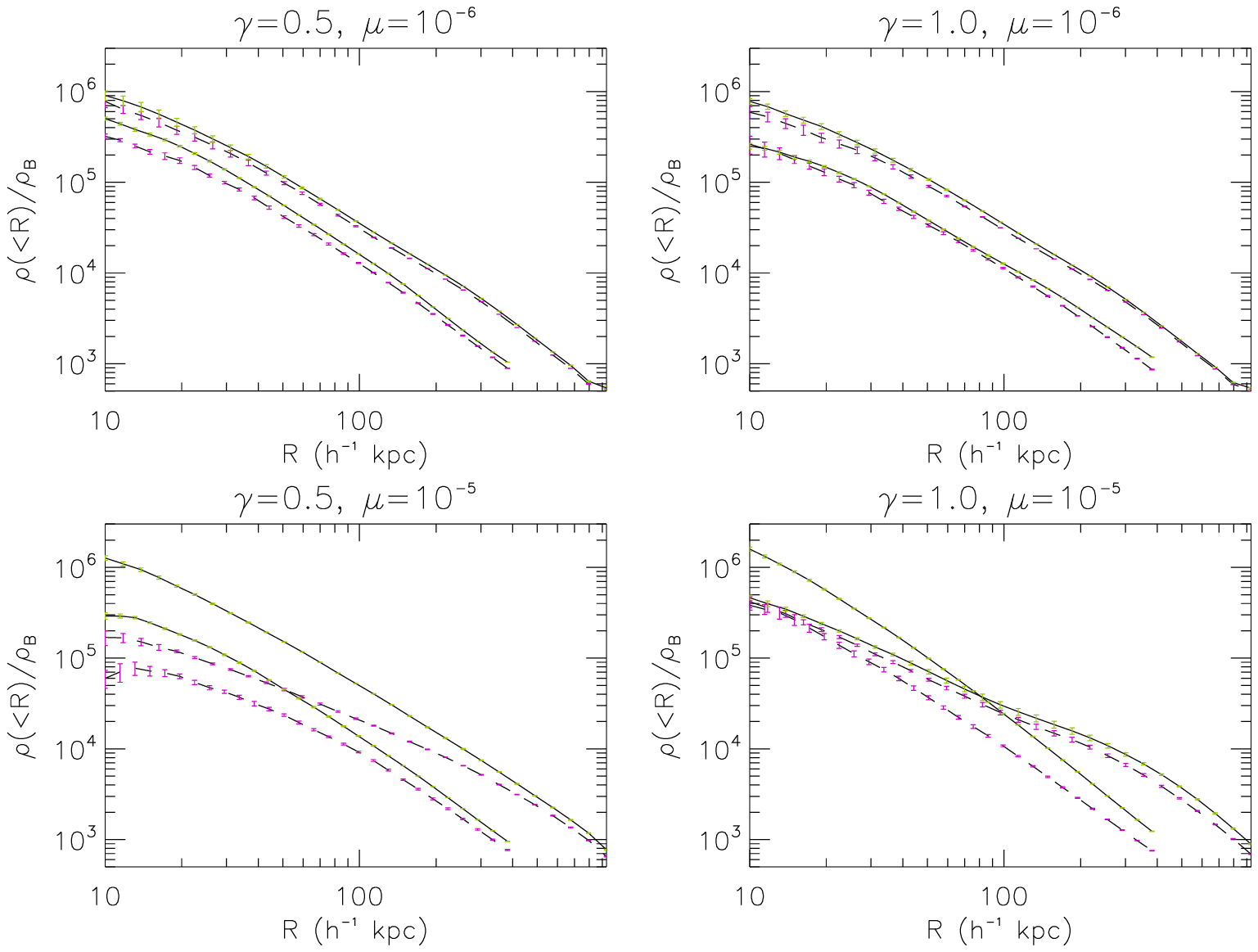}
\caption{The CDM and baryonic overdensities inside the two halos
selected above, as a function of the halo radius $R$. The solid
curves and error bars denote respectively the mean and standard
deviation for the CDM results from our 5 sample runs, while the
dashed curve and error bars are for baryons. In each panel the
upper two curves are for halo I and the lower two curves are for
halo II. $\rho_{\mathrm{B}}$ is the background matter (baryons and
CDM) density. We skipped the result for $R<10h^{-1}$~kpc, which is
not reliable due to the limit from resolution.}
\label{fig:Figure10}
\end{figure*}

We have selected 2 typical halos from each simulation. Halo I is
centred on $(x,y,z) \sim (27.0,32.2,17.8)h^{-1}$~Mpc, which is
slightly different for different simulations, and has a virial
mass $M_{vir}\sim1.16\times10^{14}h^{-1}~M_{\bigodot}$; halo II is
centred on $(x,y,z) \sim (52.5,62.9,38.3)h^{-1}$~Mpc, which is
also slightly different for different simulations, and has a
virial mass $M_{vir}\sim1.70\times10^{13}h^{-1}~M_{\bigodot}$
(note here that the virial masses are for the \emph{$\Lambda$CDM
simulations}, and for scalar field simulations they can be, and
generally are, slightly different).

The largest halos, such as halo I, generally reside in the higher
density regions where the scalar field has a heavier mass and
shows stronger chameleon effect. Consequently the fifth force
inside them is severely suppressed so that we can expect small
deviation from the $\Lambda$CDM halo profile. On the other hand,
the intermediate and small halos (such as halo II) are mostly in
relatively low density regions in which scalar field has a light
mass and the fifth force tends to saturate; this means that they
should generally have a higher internal density than the same
halos in the $\Lambda$CDM simulation due to the enhanced growth by
the fifth force.

This above analysis is qualitatively confirmed by
Fig.~\ref{fig:Figure9}, where we can see that for the cases of
$\mu=10^{-6}$ (stronger chameleon) the difference between the
predictions of the coupled scalar field models and the
$\Lambda$CDM paradigm is really small for halo I. Furthermore,
this figure also shows some new and more interesting features for
the cases of $\mu=10^{-5}$. Considering halo I in the models
$\gamma=0.5, \mu=10^{-5}$ and $\gamma=1.0, \mu=10^{-5}$ for
example: in the former case, the coupled scalar field model
produces an obviously consistent higher internal density profile
than $\Lambda$CDM, from the inner to the outer regions of the
halo; while for the latter case, the density profile of the
coupled scalar field model is lower in the inner region but higher
in the outer region!

We plan to make a detailed analysis of the complexities arising
here regarding the effects of a coupled scalar field on the
internal density profiles for halos in a future work, and in this
work we will only give a brief explanation for the new feature
observed above: Fig.~\ref{fig:Figure11} shows the distributions of
particle velocities and fifth-force-to-gravity ratio at the
positions of the particles in halo I for the two models
$\gamma=0.5, \mu=10^{-5}$ and $\gamma=1.0, \mu=10^{-5}$. As we can
see there, for the model $\gamma=1.0, \mu=10^{-5}$ the large value
of $\gamma$ makes the chameleon effect strong so that the fifth
force is generally much smaller than gravity in magnitude, while
at the same time the velocities of particles are more concentrated
towards the high end, implying a significantly higher mean speed
than $\Lambda$CDM (as we have checked numerically); as a result in
the central region of the halo the particles have higher kinetic
energy than in $\Lambda$CDM but the potential is not significantly
deeper, so that particles tend to get far away from the centre of
the halo, producing a lower inner density profile. As for the
model with $\gamma=0.5, \mu=10^{-5}$, the situation is just the
opposite: the fifth force is saturated and of the same order as
gravity due to the weak chameleon effect, so that the total
potential is greatly deeper than its $\Lambda$CDM counterpart,
while at the same time the particles are not as fast as those in
$\gamma=1.0, \mu=10^{-5}$: the consequence is that the halo
accretes more particles towards its centre.

One might also have interests in how the CDM part and the baryonic
part of the halo profile differ from each other, and this is given
in Fig.~\ref{fig:Figure10}, in which we compare the CDM and baryon
density profiles of halos I and II. Obviously the CDM density is
higher than baryons everywhere, again thanks to the boost from the
fifth force. For smaller $\mu$ and larger $\gamma$ (which help
strengthen the chameleon effect and suppress the fifth force), the
difference between the two is smaller (compare the halo I in the
models $\gamma=0.5, \mu=10^{-5}$ and $\gamma=1.0,\mu=10^{-5}$).
However, in the situations where the fifth force is already
saturated or unsuppressed (such as in halo II), increasing
$\gamma$ increases the saturated value of the fifth force, and
thus can instead magnify the difference (compare the halo II in
the models $\gamma=0.5, \mu=10^{-5}$ and
$\gamma=1.0,\mu=10^{-5}$).

The above results again show the complexity of the chameleon
scalar field model as compared to other coupled scalar field
models. We will study the environment and epoch dependence of the
halo density profiles in a upcoming work.

\section{Discussion and Conclusion}

\label{sect:con}

To summarize, in this paper we have investigated into a model
where CDM and baryons couple differently to a chameleon-like
scalar field, and performed full $N$-body simulations, by directly
solving the spatial distribution of the scalar field, to study the
nonlinear structure formation under this setup. The new complexity
introduced here compared to \cite{Li2009} is that we must
distinguish between baryons and CDM so that we know how to
calculate the force upon each particle. We do this by tagging the
initial almost homogeneously distributed particle randomly so that
80\% of all particles are tagged as CDM. We then only use CDM
particles to calculate the scalar field and only applies the fifth
scalar force on CDM particles.

The coupling function (characterized by the coupling strength
$\gamma$) and bare potential (characterized by the parameter $\mu$
which controls its steepness) of the scalar field are chosen to be
the same as in \cite{Li2009}. As discussed there, the coupling of
the scalar field to CDM particles acts on the latter a fifth
force. When $\mu$ is small and $\gamma$ is large, the chameleon
effect becomes stronger, which gives the scalar field a heavy
mass, making the fifth force short-ranged and the scalar field
dependent only on the local matter density. Other ways to
strengthen the chameleon effect includes increasing the density
and decreasing the background value of the scalar field, which
itself is equivalent to increasing the background CDM energy
density. We have displayed in Figs.~\ref{fig:Figure2},
\ref{fig:Figure4} how changing the determining factors change the
scalar field configuration and the strength of the fifth force.

We have also measured the nonlinear matter power spectrum from the
simulation results and compared them with the $\Lambda$CDM
prediction. Depending on the values of $\gamma, \mu$ as well as
the background CDM density, the former can be up to $\sim80\%$
larger than the latter. Nonetheless, when the chameleon effect is
set to be strong, the deviation gets suppressed and in particular
decreases towards small scales, showing the desirable property of
chameleon models that they evade constraints on small scales. The
bias between CDM and baryons power spectra follows the same trend.

To identify virialized halos from the simulations, we have
modified MHF, MLAPM's default halo finder, so that the calculation
of the escape velocity includes the effect of the scalar field. We
find that such a modification leads to up to a-few-percent
enhancement on the mass function compared with what is obtained
using the default MHF code, because in the latter case the escape
velocity is underestimated and some particles are incorrectly
removed from the virialized halos. We find that the mass function
in the coupled scalar field models is significantly larger than
the $\Lambda$CDM result, because of the enhanced structure growth
induced by the fifth force.

Finally, we have analyzed the internal profiles of (the same) two
halos selected from each simulation. We find that when chameleon
effect is strong and fifth force is suppressed, our result is very
close to the $\Lambda$CDM prediction; this is the case for very
large halos (which generally reside in higher density regions) and
$\mu=10^{-6}$. For large halos and $\mu=10^{-5}$, the situation is
more complicated because the competition between two effects of
the coupled scalar field, namely the speedup of the particles and
the deepening of the total attractive potential, has arrived a
critical point: if the former wins, such as in the model
$\gamma=1.0, \mu=10^{-5}$, then the inner density profile can be
lower than that in $\Lambda$CDM; if the latter wins, such as in
the model $\gamma=1.0, \mu=10^{-5}$, then we expect the opposite.
For smaller halos which locate in lower density regions, the fifth
force is not suppressed that much and causes faster growth of the
structure, so the halo is more concentrated and has a higher
internal density. Meanwhile, the bias between baryons and CDM
density profiles also increase as the fifth force becomes less
suppressed, which is as expected.

Our results already show that new features can be quantitatively
studied with the $N$-body method and the improving supercomputing
techniques, and that the chameleon model has rather different
consequences from other coupled scalar field models. The
enhancement in the structure formation due to the fifth force is
significant for some of our parameter space, which means other
observables, such as weak lensing, could place new constraints on
the model. Also one might be interested in how the halo profiles
would be like at different epoch of the cosmological evolution and
in different environments. These will be left as future work.

\begin{acknowledgments}

The work described here has been performed under the HPC-EUROPA
project, with the support of the European Community Research
Infrastructure Action under the FP8 "Structuring the European
Research Area" Programme. The $N$-body simulations are performed
on the Huygens supercomputer in the Netherlands, and the
post-precessing of data is performed on COSMOS, the UK National
Cosmology Supercomputer. We thank John Barrow, Kazuya Koyama,
Andrea Maccio and Gong-Bo Zhao for helpful discussions relevant to
this work, and Lin Jia for assistance in plotting the figures.
B.~Li is supported by a Research Fellowship in Applied Mathematics
at Queens' College, University of Cambridge and the STFC rolling
grant in DAMTP.

\end{acknowledgments}

\bigskip


\appendix

\section{Discretized Equations for Our $N$-body Simulations}

\label{appen:discret}

In the MLAPM code the partial differential equation
Eq.~(\ref{eq:INTPoisson}) is (and in our modified code
Eq.~(\ref{eq:INTphiEOM}) will also be) solved on discretized grid
points, and as such we must develop the discretized versions of
Eqs.~(\ref{eq:INTdxdtcomov} - \ref{eq:INTphiEOM}) to be
implemented into the code.

But before going on to the discretization, we need to address a
technical issue. As the potential is highly nonlinear, in the high
density regime the value of the scalar field
$\sqrt{\kappa}\varphi$ will be very close to 0, and this is
potentially a disaster as during the numerical solution process
the value of $\sqrt{\kappa}\varphi$ might easily go into the
forbidden region $\varphi<0$ \cite{Oyaizu2008}. One way of solving
this problem is to define $\chi=\bar{\chi}e^{u}$ in which
$\bar{\chi}$ is the background value of $\chi$, as in
\cite{Oyaizu2008}. Then the new variable $u$ takes value in
$(-\infty, \infty)$ so that $e^{u}$ is positive definite which
ensures that $\chi>0$. However, since there are already
exponentials of $\chi$ in the potential, this substitution will
result terms involving $\exp\left[\exp(u)\right]$, which could
potentially magnify any numerical error in $u$.

Instead, we can define a new variable $u$ according to
\begin{eqnarray}
e^{u}+1 &=& e^{\chi}.
\end{eqnarray}
By this, $u$ still takes value in $(-\infty, \infty)$,
$e^{u}\in(0, \infty)$ and thus $e^{\chi}\in(1, \infty)$ which
ensures that $\chi$ is positive definite in numerical solutions.
Besides, $e^{\beta\chi}=\left[1+e^{u}\right]^{\beta}$ so that
there will be no exponential-of-exponential terms, and the only
exponential is what we have for the potential itself. $\beta=-1$
above.

Then the Poisson equation becomes
\begin{eqnarray}\label{eq:u_Poisson}
&&\nabla^{2}\Phi_{c}\nonumber\\ &=&
\frac{3}{2}\Omega_{\mathrm{CDM}}\left[\rho_{c,\mathrm{CDM}}
\left(1+e^{u}\right)^{\gamma} -
e^{\gamma\sqrt{\kappa}\bar{\varphi}}\right]\nonumber\\
&& +
\frac{3}{2}\Omega_{\mathrm{B}}\left(\rho_{c,\mathrm{B}}-1\right) -
\frac{3\Omega_{V_{0}}a^{3}}{\left[1-\left(1+e^{u}\right)^{\beta}\right]^{\mu}}
+ 3\bar{\Omega}_{V}a^{3},\ \ \ \ \
\end{eqnarray}
where we have defined $\bar{\Omega}_{V}=\kappa
V(\bar{\varphi})/3H_{0}^{2}$ which is determined by background
cosmology, the quantity $e^{\gamma\sqrt{\kappa}\bar{\varphi}}$ is
also determined solely by background cosmology. These background
quantities should not bother us here.

The scalar field EOM becomes
\begin{eqnarray}\label{eq:u_phi_EOM}
&&\frac{ac^{2}}{\left(H_{0}B\right)^{2}}\nabla\cdot\left(\frac{e^{u}}{1+e^{u}}\nabla
u\right)\nonumber\\
&=& 3\gamma\Omega_{\mathrm{CDM}}\rho_{c,\mathrm{CDM}}
\left(1+e^{u}\right)^{\gamma} +
\frac{3\mu\beta\Omega_{V_{0}}a^{3}\left(1+e^{u}\right)^{\beta}}
{\left[1-\left(1+e^{u}\right)^{\beta}\right]^{\mu+1}}\nonumber\\
&& -
3\gamma\Omega_{\mathrm{CDM}}e^{\gamma\sqrt{\kappa}\bar{\varphi}} -
\frac{3\mu\beta\Omega_{V_{0}}a^{3}e^{\beta\sqrt{\kappa}\bar{\varphi}}}
{\left[1-e^{\beta\sqrt{\kappa}\bar{\varphi}}\right]^{m+1}}
\end{eqnarray}
in which we have used the fact that
$\chi=\log(1+e^{u})\Rightarrow\nabla\chi=\frac{e^{u}}{1+e^{u}}\nabla
u$, and moved all terms depending only on background cosmology
(the source terms) to the right hand side.

So, in terms of the new variable $u$, the set of equations used in
the $N$-body code should be
\begin{eqnarray}\label{eq:u_dxdt}
\frac{d\mathbf{x}_{c}}{dt_{c}} &=& \frac{\mathbf{p}_{c}}{a^{2}},\\
\label{eq:u_dpdt} \frac{d\mathbf{p}_{c}}{dt_{c}} &=&
-\frac{1}{a}\nabla\Phi_{c} \left[-
\frac{c^{2}\gamma}{\left(H_{0}B\right)^{2}}\frac{e^{u}}{1+e^{u}}\nabla
u\right]
\end{eqnarray}
plus Eqs.~(\ref{eq:u_Poisson}, \ref{eq:u_phi_EOM}). These
equations will ultimately be used in the code. Among them,
Eqs.~(\ref{eq:u_Poisson}, \ref{eq:u_dpdt}) will use the value of
$u$ while Eq.~(\ref{eq:u_phi_EOM}) solves for $u$. In order that
these equations can be integrated into MLAPM, we need to
discretize Eq.~(\ref{eq:u_phi_EOM}) for the application of
Newton-Gauss-Seidel iterations.

To discretize Eq.~(\ref{eq:u_phi_EOM}), let us define
$b\equiv\frac{e^{u}}{1+e^{u}}$. The discretization involves
writing down a discretion version of this equation on a uniform
grid with grid spacing $h$. Suppose we require second order
precision as is in the standard Poisson solver of MLAPM, then
$\nabla u$ in one dimension can be written as
\begin{eqnarray}
\nabla u &\rightarrow& \nabla^{h}u_{j}\ =\
\frac{u_{j+1}-u_{j-1}}{2h}
\end{eqnarray}
where a subscript $_{j}$ means that the quantity is evaluated on
the $j$-th point. Of course the generalization to three dimensions
is straightforward.

The factor $b$ in $\nabla\cdot\left(b\nabla u\right)$ makes this a
standard variable coefficient problem. We need also discretize
$b$, and do it in this way (again for one dimension):
\begin{widetext}
\begin{eqnarray}
\nabla\cdot\left(b\nabla u\right) &\rightarrow&
\left(\nabla^{h}b_{j}\right)\cdot\left(\nabla^{h}u_{j}\right) +
b_{j}\nabla^{h2}u_{j}\nonumber\\
&=& \frac{b_{j+1/2}-b_{j-1/2}}{h}\frac{u_{j+1}-u_{j-1}}{2h} +
\frac{b_{j+1/2}+b_{j-1/2}}{2}\frac{u_{j+1}-2u_{j}+u_{j-1}}{h^{2}}\nonumber\\
&=& \frac{1}{h^{2}}\left[b_{j+\frac{1}{2}}u_{j+1} -
u_{j}\left(b_{j+\frac{1}{2}}+b_{j-\frac{1}{2}}\right) +
b_{j-\frac{1}{2}}u_{j-1}\right]
\end{eqnarray}
\end{widetext}
where we have defined
$b_{j+\frac{1}{2}}=\left(b_{j}+b_{j+1}\right)/2$ and
$b_{j-\frac{1}{2}}=\left(b_{j-1}+b_{j}\right)/2$. This can be
easily generalize to three dimensions as
\begin{widetext}
\begin{eqnarray}
\nabla\cdot\left(b\nabla u\right) &\rightarrow&
\frac{1}{h^{2}}\left[b_{i+\frac{1}{2},j,k}u_{i+1,j,k} -
u_{i,j,k}\left(b_{i+\frac{1}{2},j,k}+b_{i-\frac{1}{2},j,k}\right)
+ b_{i-\frac{1}{2},j,k}u_{i-1,j,k}\right]\nonumber\\
&& + \frac{1}{h^{2}}\left[b_{i,j+\frac{1}{2},k}u_{i,j+1,k} -
u_{i,j,k}\left(b_{i,j+\frac{1}{2},k}+b_{i,j-\frac{1}{2},k}\right)
+ b_{i,j-\frac{1}{2},k}u_{i,j-1,k}\right]\nonumber\\
&& + \frac{1}{h^{2}}\left[b_{i,j,k+\frac{1}{2}}u_{i,j,k+1} -
u_{i,j,k}\left(b_{i,j,k+\frac{1}{2}}+b_{i,j,k-\frac{1}{2}}\right)
+ b_{i,j,k-\frac{1}{2}}u_{i,j,k-1}\right].
\end{eqnarray}
\end{widetext}
Then the discrete version of Eq.~(\ref{eq:u_phi_EOM}) is
\begin{eqnarray}\label{eq:diffop}
L^{h}\left(u_{i,j,k}\right) &=& 0,
\end{eqnarray}
in which
\begin{widetext}
\begin{eqnarray}
L^{h}\left(u_{i,j,k}\right) &=&
\frac{1}{h^{2}}\left[b_{i+\frac{1}{2},j,k}u_{i+1,j,k} -
u_{i,j,k}\left(b_{i+\frac{1}{2},j,k}+b_{i-\frac{1}{2},j,k}\right)
+ b_{i-\frac{1}{2},j,k}u_{i-1,j,k}\right]\nonumber\\
&& + \frac{1}{h^{2}}\left[b_{i,j+\frac{1}{2},k}u_{i,j+1,k} -
u_{i,j,k}\left(b_{i,j+\frac{1}{2},k}+b_{i,j-\frac{1}{2},k}\right)
+ b_{i,j-\frac{1}{2},k}u_{i,j-1,k}\right]\nonumber\\
&& + \frac{1}{h^{2}}\left[b_{i,j,k+\frac{1}{2}}u_{i,j,k+1} -
u_{i,j,k}\left(b_{i,j,k+\frac{1}{2}}+b_{i,j,k-\frac{1}{2}}\right)
+ b_{i,j,k-\frac{1}{2}}u_{i,j,k-1}\right]\nonumber\\
&&-\frac{\left(H_{0}B\right)^{2}}{ac^{2}}\left[3\gamma\Omega_{\mathrm{CDM}}\rho^{\mathrm{CDM}}_{c,i,j,k}
\left(1+e^{u_{i,j,k}}\right)^{\gamma} +
\frac{3\mu\beta\Omega_{V_{0}}a^{3}\left(1+e^{u_{i,j,k}}\right)^{\beta}}
{\left[1-\left(1+e^{u_{i,j,k}}\right)^{\beta}\right]^{\mu+1}}\right]\nonumber\\
&& +\frac{\left(H_{0}B\right)^{2}}{ac^{2}}\left[
3\gamma\Omega_{\mathrm{CDM}}e^{\gamma\sqrt{\kappa}\bar{\varphi}} +
\frac{3\mu\beta\Omega_{V_{0}}a^{3}e^{\beta\sqrt{\kappa}\bar{\varphi}}}
{\left[1-e^{\beta\sqrt{\kappa}\bar{\varphi}}\right]^{\mu+1}}\right].
\end{eqnarray}
\end{widetext}
Then the Newton-Gauss-Seidel iteration says that we can obtain a
new (and often more accurate) solution of $u$,
$u^{\mathrm{new}}_{i,j,k}$, using our knowledge about the old (and
less accurate) solution $u^{\mathrm{old}}_{i,j,k}$ as
\begin{eqnarray}\label{eq:GS}
u^{\mathrm{new}}_{i,j,k} &=& u^{\mathrm{old}}_{i,j,k} -
\frac{L^{h}\left(u^{\mathrm{old}}_{i,j,k}\right)}{\partial
L^{h}\left(u^{\mathrm{old}}_{i,j,k}\right)/\partial u_{i,j,k}}.
\end{eqnarray}
The old solution will be replaced by the new solution to
$u_{i,j,k}$ once the new solution is ready, using the red-black
Gauss-Seidel sweeping scheme. Note that
\begin{widetext}
\begin{eqnarray}
\frac{\partial L^{h}(u_{i,j,k})}{\partial u_{i,j,k}} &=&
\frac{1}{2h^{2}}\frac{e^{u_{i,j,k}}}{\left(1
+e^{u_{i,j,k}}\right)^{2}}\left[u_{i+1,j,k}+u_{i-1,j,k}+u_{i,j+1,k}
+u_{i,j-1,k}+u_{i,j,k+1}+u_{i,j,k-1}-6u_{i,j,k}\right]\nonumber\\
&&-\frac{1}{2h^{2}}\left[b_{i+1,j,k}+b_{i-1,j,k}+b_{i,j+1,k}
+b_{i,j-1,k}+b_{i,j,k+1}+b_{i,j,k-1}+6b_{i,j,k}\right]\nonumber\\
&&-\frac{\left(H_{0}B\right)^{2}}{ac^{2}}3\gamma^{2}\Omega_{\mathrm{CDM}}\rho^{\mathrm{CDM}}_{c,i,j,k}
\left(1+e^{u_{i,j,k}}\right)^{\gamma}b_{i,j,k}\nonumber\\
&&-\frac{\left(H_{0}B\right)^{2}}{ac^{2}}
\frac{3\mu\beta^{2}\Omega_{V_{0}}a^{3}\left(1+e^{u_{i,j,k}}\right)^{\beta}}
{\left[1-\left(1+e^{u_{i,j,k}}\right)^{\beta}\right]^{\mu+1}}b_{i,j,k}
\left[1+(\mu+1)\frac{\left(1+e^{u_{i,j,k}}\right)^{\beta}}{1-\left(1+e^{u_{i,j,k}}\right)^{\beta}}\right].
\end{eqnarray}
\end{widetext}

In principle, if we start from a high redshift, then the initial
guess of $u_{i,j,k}$ could be such that the initial value of
$\chi$ in all the space is equal to the background value
$\bar{\chi}$, because anyway at this time we expect this to be
approximately true. For subsequent time steps we could use the
solution for $u_{i,j,k}$ at the previous time step as our initial
guess; if the time step is small enough then we do not expect $u$
to change significantly between consecutive times so that such a
guess will be good enough for the iteration to converge fast.

In practice, however, due to specific features and algorithm of
the MLAPM code \cite{MLAPM}, the above procedure may be slightly
different in details.

\end{document}